\documentclass[sigconf, screen, nonacm]{acmart}


\usepackage{tikz}
\usepackage{pgfplots}
\usepackage{subcaption}
\pgfplotsset{compat=1.18,
every axis/.append style={axis lines=left,axis line style={black,line width=.45pt},
tick align=outside,major tick length=2pt,minor tick num=0,
xmajorgrids=false,ymajorgrids=false,xminorgrids=false,yminorgrids=false,
every axis plot/.append style={line join=round}}}
\usetikzlibrary{positioning,arrows.meta,calc,fit,backgrounds,shapes.misc,patterns}
\usepackage{booktabs}
\usepackage{tabularx}
\usepackage{multirow}
\usepackage{ragged2e}
\usepackage{array}
\usepackage{xspace}

\newcommand{\stalllens}{\textsc{StallLens}\xspace}
\newcommand{\dmem}{Design Memory\xspace}
\newcounter{obs}

\definecolor{ink}{HTML}{000000}
\definecolor{slate}{HTML}{333333}
\definecolor{steel}{HTML}{707070}
\definecolor{mist}{HTML}{D9D9D9}
\definecolor{fog}{HTML}{FFFFFF}
\definecolor{accent}{HTML}{000000}

\tikzset{
  box/.style={draw=black,fill=white,sharp corners,align=center,font=\footnotesize,inner sep=3pt},
  dbox/.style={draw=black,fill=white,sharp corners,align=center,font=\footnotesize,inner sep=3pt},
  hbox/.style={draw=accent,fill=white,sharp corners,align=center,font=\footnotesize,inner sep=3pt},
  arr/.style={-{Stealth[length=4pt]},semithick,draw=slate},
  lab/.style={font=\scriptsize,text=ink,fill=white,inner sep=1pt},
}

\usepackage{xurl,placeins}
\usepackage{algorithm,algpseudocode}
\begin{document}
\title[DELTA: Diagnosing Exploration Lineages via Twin-path Alignment]{DELTA: Diagnosing Exploration Lineages via Twin-path Alignment — What Can Contrasting Design Paths Reveal About Microarchitectural Behavior?}

\author{Kaushal Mhapsekar}
\affiliation{%
  \institution{North Carolina State University}
  \city{Raleigh}
  \state{North Carolina}
  \country{USA}}
\email{kmhapse@ncsu.edu}

\author{Darsh Asher}
\affiliation{%
  \institution{North Carolina State University}
  \city{Raleigh}
  \state{North Carolina}
  \country{USA}}
\email{dkasher@ncsu.edu}

\author{Bita Aslrousta}
\affiliation{%
  \institution{North Carolina State University}
  \city{Raleigh}
  \state{North Carolina}
  \country{USA}}
\email{baslrou@ncsu.edu}

\author{Amna Hassan}
\affiliation{%
  \institution{North Carolina State University}
  \city{Raleigh}
  \state{North Carolina}
  \country{USA}}
\email{ahassan9@ncsu.edu}

\author{Debayan Sengupta}
\affiliation{%
  \institution{North Carolina State University}
  \city{Raleigh}
  \state{North Carolina}
  \country{USA}}
\email{dsengup2@ncsu.edu}

\author{Azam Ghanbari}
\affiliation{%
  \institution{North Carolina State University}
  \city{Raleigh}
  \state{North Carolina}
  \country{USA}}
\email{aghanba2@ncsu.edu}

\author{Samira Mirbagher Ajorpaz}
\affiliation{%
  \institution{North Carolina State University}
  \city{Raleigh}
  \state{North Carolina}
  \country{USA}}
\email{smirbag@ncsu.edu}

\renewcommand{\shortauthors}{Mhapsekar et al.}

\begin{abstract}
Architects have two mature instruments for performance work. Top-Down analysis tells them where a design loses its pipeline slots, and design-space exploration (DSE) tells them which configuration scores best. Neither explains why a search stops improving: why a backend-bound design rejects a prefetcher that the bottleneck label recommends, or why a capacity increase that helped in one configuration hurts in the next. This paper argues that the evidence needed to answer such questions is already produced during exploration and then discarded. Every parent-child edge in an exploration tree is a controlled intervention: one knob changes, the rest of the design is held fixed, and a cycle-level simulator measures the outcome. When the same edit helps under one parent and hurts under another, aligning the two parents localizes the interaction that decides its sign.

We present DELTA, a simulator-grounded root-cause analysis tool that builds a Design Memory of such interventions during exploration and lets an architect query it afterwards. Each record stores the full parent configuration, the single applied edit, the measured and predicted IPC, a Top-Down breakdown from StallLens, our gem5 implementation of Top-Down, and its lineage. Unlike evolutionary LLM search, which keeps only high-scoring programs, DELTA keeps failed edges as evidence. The same memory also makes exploration efficient: on four SPEC CPU2017 workloads, the full system raises IPC by 55.5\% in 22 simulated designs for \$1.56 of model cost, predicts the IPC of its own proposals to within 3.3\%, and keeps improving after a bottleneck-guided search stalls. A root-cause query over the recorded history led us to a usefulness-throttled variant of the Best-Offset prefetcher that improves IPC by 1.37\% over the original in matched configurations.

\end{abstract}
\maketitle
\pagestyle{plain}

\suppressfloats[t]
\begin{figure}[t]
\centering
\captionsetup{skip=2pt}
\begin{tikzpicture}
\begin{axis}[
width=.98\linewidth,height=28mm,
xmin=0,xmax=25,ymin=.94,ymax=1.55,
xtick={0,5,10,15,20,25},ytick={1,1.25,1.5},
xlabel={Design index},ylabel={Best IPC so far},
label style={font=\footnotesize},tick label style={font=\footnotesize},
legend style={font=\footnotesize,draw=none,fill=none,
at={(.5,1.03)},anchor=south,legend columns=2},
axis line style={black,line width=.45pt},grid=none]
\addplot[black,solid,semithick,const plot,mark=*,mark size=1.1pt,mark repeat=5]
coordinates {(0,0.96350) (1,0.99350) (2,1.31420) (3,1.31420) (4,1.31420) (5,1.32830) (6,1.32830) (7,1.38520) (8,1.38520) (9,1.41180) (10,1.41180) (11,1.41180) (12,1.41950) (13,1.44480) (14,1.44630) (15,1.44630) (16,1.45240) (17,1.45410) (18,1.45410) (19,1.46100) (20,1.49620) (21,1.49780) (22,1.49780) (23,1.49780) (24,1.49780) (25,1.49780)};
\addlegendentry{DELTA-Full}
\addplot[black,dashed,semithick,const plot,mark=square,mark size=1.3pt,
mark repeat=5,mark options={solid,fill=white}]
coordinates {(0,0.96350) (1,0.96420) (2,1.20650) (3,1.20650) (4,1.20650) (5,1.20650) (6,1.23380) (7,1.23380) (8,1.24890) (9,1.24890) (10,1.24890) (11,1.25840) (12,1.27690) (13,1.27690) (14,1.27690) (15,1.27750) (16,1.27750) (17,1.28030) (18,1.28030) (19,1.28110) (20,1.28530) (21,1.29440) (22,1.30120) (23,1.34540) (24,1.34540) (25,1.34540)};
\addlegendentry{DELTA-Guide}
\end{axis}
\end{tikzpicture}
\caption{Planning sustains gains over 25 edits: Full plans alternatives; Guide uses one proposal with Top-Down feedback.}
\Description{Best recorded IPC versus design index for DELTA-Full and DELTA-Guide. Full uses planning and bottleneck feedback; Guide uses bottleneck feedback with one proposal. Solid and dashed step curves distinguish their recorded search histories.}
\label{fig:plateau}
\end{figure}
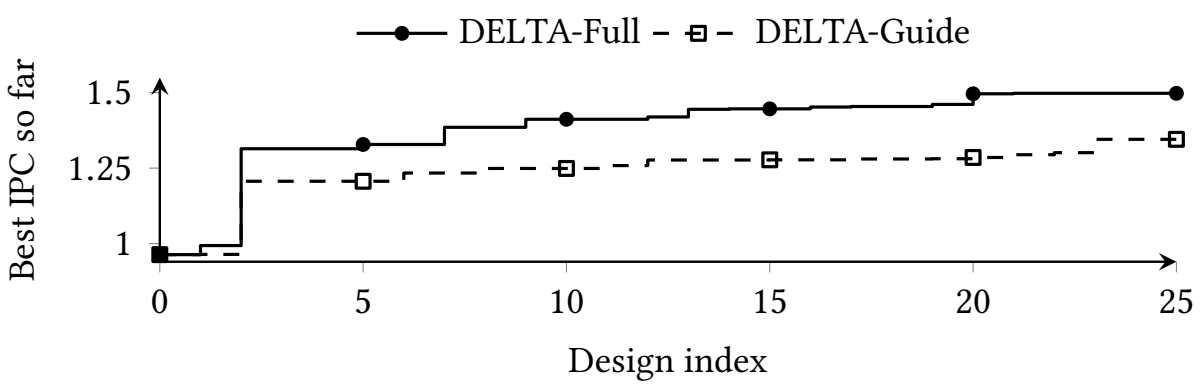

\section{Introduction}
\label{sec:intro}

Improving a processor starts with identifying the root cause of each performance bottleneck: which structure must change to remove the stall and raise instructions per cycle (IPC)? The answer is rarely a property of one structure. 
Cache misses, branch speculation, and instruction window occupancy overlap, so the benefit of a larger cache, a different predictor, or a deeper reorder buffer depends on the processor being changed. 
Microarchitectures and applications also evolve, changing which interactions limit performance~\cite{archgym_isca23,concorde_isca25}. Architects must therefore locate and explain the interactions that limit each new design, not only find a fast configuration.

Consider one change measured on two processors. A prefetcher requests a predicted future cache line before the program asks for it. Enabling one at the second cache level (L2) leaves IPC unchanged ($-0.01\%$) with a 32\,KB first-level instruction cache (L1I) and raises it by 20.21\% once that cache holds 64\,KB. The prefetcher is identical in both cases; what differs is how much work the frontend supplies. 

Common practice follows the largest stall category reported by Top-Down analysis~\cite{yasin_ispass14}, an accounting of where pipeline slots are lost, and enlarges the resource it names. This gives early gains but then plateaus: mean IPC gain rises from 33\% at ten iterations to only 35\% at twenty iterations; DELTA reaches 46\% at ten (Section~\ref{sec:search}). Figure~\ref{fig:plateau} contrasts the complete recorded paths with and without planning. Aggressive capacity growth also stalls: doubling the L1 data cache (L1D) from 64 to 128\,KB gains 6.8\%, but doubling again to 256\,KB loses 0.30\%. L1D filters requests reaching the L2 prefetcher and changes what it learns. Further progress requires resolving this interaction, rather than continuing to enlarge the resource named by the stall category (Section~\ref{sec:abop}). Four problems motivate this diagnosis.

\paragraph{Search finds fast designs without explaining them.}
\begingroup\setlength{\emergencystretch}{3em}
Design-space exploration treats the processor as a black box to be optimized. ArchRanker~\cite{archranker_isca14} ranks candidate pairs to steer search, BOOM-Explorer~\cite{boomexplorer_iccad21} combines microarchitecture-aware active learning with Bayesian optimization, ArchExplorer~\cite{archexplorer_micro23} shifts resources toward active bottlenecks, and ArchGym~\cite{archgym_isca23} connects search agents to simulators through a common interface. The evaluations behind a selected configuration are expensive: in our setup, one evaluation costs roughly 120 core hours across 40 SimPoint slices~\cite{simpoint_asplos02}. A best design alone does not explain its interactions or which conditional responses to test again when the workload or baseline changes.
\par\endgroup

\paragraph{Learned performance models require labels.}
\noindent Supervised neural predictors~\cite{ipek_asplos06} and analytical models combined with learning~\cite{concorde_isca25} reduce the cost of evaluating a configuration, but they require labels for the design space they predict: measured results paired with model inputs, such as the IPC of a particular configuration and workload. Constructing a labeled design space requires sustained simulation effort. As hardware and workloads change, maintaining that coverage requires additional measurements: changes to the core, predictor family, or benchmark suite introduce new contexts whose performance response must be represented in the predictor's inputs or adaptation process. The labels also answer a narrower question than the architect asks. A model that predicts the IPC of a configuration does not state which change caused the difference from its predecessor or under which conditions that change helps. We seek an assistant that uses new observations directly, without another training dataset.

\paragraph{Counters do not identify the root cause.}
\noindent An issue slot is an opportunity to send an operation for execution. A stall leaves such progress blocked; wrong-path work executes after an incorrect branch prediction and must be discarded when the branch resolves.
Bottleneck analysis explains a single execution. Top-Down analysis~\cite{yasin_ispass14} attributes lost issue slots to frontend, backend, and speculation categories; critical-path prediction~\cite{fields_isca01} identifies the instructions that determine execution time; interaction cost analysis~\cite{fields_micro03} measures how the latency of overlapping events differs from the sum of their individual latencies; and Gus~\cite{gus24} perturbs modeled resources to find instruction-level timing sensitivities. These methods show where time goes in one configuration, but a large stall fraction does not identify the change that will reduce it. 

Because slot fractions sum to one, relieving one stall can enlarge another category's share even as IPC rises: enabling the SlimAMPM prefetcher~\cite{slimampm_dpc15} on one parent raises IPC by 32.28\% while the Bad Speculation share (slots spent on wrong-path work) rises from 14.00\% to 19.84\%. The reverse also occurs. When the workload is a transformer's forward pass, a prefetch controller that skips each page's prefix raises prefetch accuracy from 0.035653 to 0.983862 and cuts L1 demand misses by 46.50\%, yet execution takes 61.72\% more cycles because the skipped lines return as new demand misses. The first contrast shows a larger stall share with higher IPC; the second shows better prefetch accuracy and fewer L1 misses with longer execution. Neither counter direction alone determines the performance response. The required comparison is counterfactual: it places a processor (the parent) beside the same processor after one change (its child).

\paragraph{Search scores and diagnostic records serve different purposes.}
\noindent Search with language models now both proposes and evaluates changes. 
FunSearch~\cite{funsearch_nature24} and AlphaEvolve~\cite{alphaevolve25} pair programs generated by the model with an automated evaluator and keep the candidates that score best. Selecting candidates by score does not itself explain why a change helps one parent and hurts another; that explanation requires neutral and regressive outcomes linked to checked parent contexts. The same edit can have different effects across parents. Enlarging the reorder buffer (ROB) from 192 to 256 entries gains 4.28\% with the TAGE branch predictor, which matches tagged tables of increasing branch history length, but 0.90\% with TAGE-SC-L~\cite{seznec_cbp16}, which adds statistical correction and loop prediction to TAGE; these four configurations hold other settings fixed, isolating the predictor's effect on ROB gains. This predictor is hardware that speculates on control flow, not the model that predicts IPC. Repeated proposals make retaining these comparisons especially valuable. In our runs, 12 TAGE-SC-L proposals follow an earlier regression in the same run; ten of them lower IPC again, and the 12 consume roughly 1{,}440 core hours. DELTA retains these outcomes with their parents, turning the simulation effort into a record of the contexts in which the predictor reduces throughput. The resulting comparisons connect proposal selection to architectural diagnosis.

\paragraph{Gaps in current methods.}
These problems motivate a common diagnostic record that connects search history, performance prediction, and bottleneck analysis. Configuration scores describe the quality of a design; parent--child links explain the change that produced it. Learned predictors organize measured labels, while bottleneck analysis attributes execution time to resources and events. Aligning these objects across contexts makes the response to the same edit an explicit subject of analysis. CacheMind~\cite{cachemind_asplos26} grounds cache explanations in execution traces, but it answers questions about one structure within a run rather than aligning measured changes across processor configurations. Retaining evaluations during search preserves outcomes; diagnosis additionally requires checks that align edits and parent contexts. Each child must be tied to a checked parent configuration, an identified edit, fixed workload weights, and equal measurement conditions, so that two records of the same edit can be compared. Genetic background analysis~\cite{chandler_tig13} offers a model for this comparison: it tests one mutation across genetic backgrounds to reveal which background modifies its effect.

Architects already learn through such trials: propose, measure, compare, and revise. An assistant built on a pretrained language model should retain these sequences, beyond configurations and scores. Retrieval-augmented generation supplies selected stored records as part of that model's input; in-context learning uses those examples to interpret a new query without changing model weights. These operations let it reuse observations, suggest broader relationships, and test their conditions. The model organizes evidence and proposes tests; the architect examines explanations and directs changes.

DELTA (Diagnosing Exploration Lineages via Twin-path Alignment) turns the history of a design search into diagnostic evidence. It keeps every simulated child beside its parent, including neutral and regressive children, retrieves linked trials, predicts child IPC, and proposes changes for simulation. A pretrained language model with fixed weights reads the retrieved records and returns each proposal; the simulator alone determines its outcome. Evidence updates memory and context, without task-specific supervised training, backpropagation, or fine-tuning. DELTA first checks that every recorded path begins at the stated baseline configuration, that repeated parent measurements agree, and that the requested edit changes the simulated configuration. It then compares full parent configuration vectors to identify candidate interactions. Four matched configurations isolate an edit's interaction with a context parameter. Workload and slot deltas connect each performance response to the affected execution behavior and direct tests that distinguish mechanisms. Evidence levels identify the measurements supporting each explanation, from an observed IPC change through a controlled intervention and its repetition in further contexts.

We evaluate 100 linked interventions across four variants and four SPEC CPU2017 workloads in gem5~\cite{gem5_v20}. Planning improves peak IPC; bottleneck feedback speeds convergence. At ten iterations, DELTA raises mean IPC by 46\%. Controlled width transfer, from a four-wide to an eight-wide core, reaches 0.87\% prediction error with 16 target measurements and no weight updates. Changes to branch prediction and prefetching account for 18 of the 19 retained edits that lower IPC; faster memory service can expose other stalls despite higher IPC. Like genetic suppression~\cite{matsui_bioessays17}, which tests compensation by a second mutation, DELTA tests repairs by comparing IPC in matched configurations. Aligned L1D paths expose changes in Best-Offset prefetching's (BOP)~\cite{bop_hpca16} L2 request stream. BOP prefetches at the address offset that scores best on recent requests. Adaptive BOP keeps this learning and adds two controls directed by feedback~\cite{fdp_hpca07}: usefulness control issues prefetches according to the ratio of useful to issued ones, and retraining clears the learned scores when the request stream changes, gaining 1.37\% mean IPC.

Targeted tests connect performance gains to recovery throughput, memory capacity, and prefetch coverage. Squash width is the number of wrong-path instructions the reorder buffer can discard per cycle after a branch misprediction. Increasing squash width from 8 to 32 gains 29.38\% with LocalBP, gem5's predictor that uses each branch's own outcome history, and 21.03\% with TAGE on random branches; width 64 raises these gains to 37.95\% and 25.67\%. Conditional admission, which limits new instructions entering the backend when recovery consumes many cycles, gains 5.66\% on the LocalBP case. Miss status holding registers (MSHRs) track cache lines with requests still outstanding; accesses to the same line can share an entry. Doubling physical MSHRs from two to four gains 99.19--99.87\% across three contended load cases.

For a transformer workload, executed by the simulated processor and separate from the model that proposes edits, page reset and interval filling improve prefetch coverage, gaining 60.8\% and 55.3\% over stride prefetching at prefix positions 16 and 128. A prefix position identifies how many preceding tokens supply the saved context. The repair resets prefetch progress at a memory page boundary and requests intervening lines that a distant target would skip. Shorter execution reduces modeled core/cache energy by 16.89\% and 15.91\% at prefix positions 16 and 128: the reduced execution time more than offsets the increase in modeled power. These cases connect each measured gain to a change in the limiting resource. DELTA is also portable to new architectures without supervised learning or retraining: applied unchanged to RISC-V cores, elimination follows the same steps to the same request supply cause on the smaller core, and the unchanged repair removes every candidate crossing a page.

This paper makes the following contributions:
\begin{itemize}
\item \textbf{Diagnosis from parent--child deltas.} DELTA retains every simulated parent--child step, checks edge consistency, and aligns repeated edits across parents to measure context effects and pairwise interactions (Section~\ref{sec:design}).
\item \textbf{IPC prediction without offline training.} The proposing model predicts child IPC from the active run's history with 3.30\% mean absolute percentage error (1.33\% median) in Guide (with bottleneck feedback) against 8.24\% in Record (without it), and Guide's error falls from 4.96\% to 1.49\% between run halves. Controlled width transfer reaches 0.87\% error through measurement memory without weight updates (Sections~\ref{sec:surrogate} and~\ref{sec:transfer}).
\item \textbf{A classification of hardware interventions.} DELTA separates edit semantics, parent context, outcome, mechanism, and evidence level, and applies them to 100 interventions to extract parameter scaling rules (Section~\ref{sec:taxonomy}).
\item \textbf{Slot delta accounting and an intervention dataset.} \stalllens, DELTA's Top-Down slot accounting, pairs each child's slot deltas with its IPC delta inside the search loop and matches the dominant hardware category in 17 of 22 workloads. The dataset contains the 100-edge store and prompts.
\item \textbf{Search and prefetch control.} At ten iterations DELTA raises IPC by 46\%, compared with 36\%, 33\%, and 19\% for genetic, Top-Down greedy, and random search. Its four variants gain 32.93--55.46\%; Adaptive BOP derived from L1D path alignment gains 1.37\% mean IPC (Sections~\ref{sec:search} and~\ref{sec:abop}).
\item \textbf{Recovery and memory case studies.} Raising squash width, the wrong-path instructions discarded per cycle, from 8 to 32 gains 29.38\%/21.03\% with the LocalBP/TAGE branch predictors, reaching 37.95\%/25.67\% at width 64. Branch admission, which narrows rename while branches are unresolved, and conditional admission gain 12.86\% and 5.66\% on the LocalBP case. Doubling physical MSHRs from two to four gains 99.19--99.87\% on three contended load cases (Section~\ref{sec:recovery}).
\item \textbf{Transformer coverage and energy.} Page reset and interval filling gain 60.8\%/55.3\% over stride at prefix positions 16/128, or $7.418\times$/$7.092\times$ over no prefetch. Modeled core/cache energy falls 16.89\%/15.91\% relative to stride as shorter execution offsets higher modeled power (Section~\ref{sec:repair}).
\item \textbf{Portability to new architectures.} Without supervised learning or retraining, the unchanged procedure and repair carry to RISC-V cores sized after SonicBOOM, an open-source out-of-order processor. DELTA's root cause analysis and bottleneck elimination follow the same steps to the same page boundary cause; the repair gains 4.90\% and 4.35\% over the distant target alone and up to 22.80\% over no prefetch. On graph-based vector search (HNSW), it rules out window, request, and cache capacity and locates loads at $\mathit{base}+396\cdot\mathit{id}$, a scale that power-of-two matching does not learn; matching by division gains 7.32\% against 0.66\% (Sections~\ref{sec:repair} and~\ref{sec:riscv}).
\end{itemize}

Sections~\ref{sec:design} and~\ref{sec:methodology} describe DELTA and the experiments. Sections~\ref{sec:results} and~\ref{sec:search} report the results, Section~\ref{sec:background} relates prior work, and Section~\ref{sec:conclusion} summarizes the findings.

\section{DELTA: Comparing Design Changes}
\label{sec:design}
\begin{figure}[t]
\centering
\begin{tikzpicture}[x=1mm,y=1mm,font=\footnotesize,
b/.style={draw=black,fill=white,align=center,font=\footnotesize,inner sep=1.6pt},
e/.style={-{Stealth[length=1.3mm]},line width=.4pt},
tw/.style={-{Stealth[length=1.5mm]},line width=1.1pt},
l/.style={font=\scriptsize,inner sep=.6pt}]
\node[font=\footnotesize\bfseries,anchor=west] at (-1.5,4.5) {(a) Exploration tree};
\node[font=\footnotesize\bfseries,anchor=west] at (48.5,4.5) {(b) Diagnosis};
\node[b] (r) at (5.5,-2) {baseline};
\node[b] (c0) at (28,-2) {$c_0$: TAGE, ROB 192};
\node[b] (k0) at (8,-13) {ROB 256};
\node[b] (c1) at (31,-13) {$c_1$: TAGE-SC-L, ROB 192};
\node[b] (k1) at (31,-22.5) {ROB 256};
\draw[e,dashed] (r)--(c0);
\draw[tw] (c0.south)--(k0.north);
\node[l,anchor=east] at (15.5,-7.4) {$E$: $+4.28\%$};
\draw[e] (c0.south)--(c1.north);
\node[l,anchor=west] at (30.5,-7.4) {$X$: $-0.17\%$};
\draw[tw] (c1)--(k1);
\node[l,anchor=west] at (32.5,-17.8) {$E$: $+0.90\%$};
\node[anchor=north west,align=left,font=\scriptsize,inner sep=0pt] at (-1,-17) {thick: twin paths, the\\same edit $E$ from two parents};
\node[b,align=left,anchor=north west,text width=32mm,font=\scriptsize] (d) at (49,2) {Align: $c_0$ and $c_1$ differ only in the predictor, $X$.\\[1pt]Compare the gain of $E$:\\$0.90-4.28=-3.38$ points.\\[1pt]Check counters; test a control.\\[1pt]\textbf{The predictor limits the \mbox{window's} return.}};
\node[b,text width=82mm] (m) at (41,-31) {\textbf{\dmem}: every parent, edit, and child with predicted and measured IPC and counters, including neutral and regressive edits};
\draw[e] (k1.south)--(k1.south |- m.north);
\node[l,anchor=west] at (32.5,-25.6) {write};
\draw[e,dashed] (d.south |- m.north)--(d.south);
\node[l,anchor=west] at (67.5,-25) {retrieve};
\end{tikzpicture}
\caption{A recorded tree (Full): one ROB edit under two predictors forms a twin comparison.}
\Description{A tree on the left: a baseline leads to a parent with TAGE and a 192-entry reorder buffer. One child enlarges the reorder buffer to 256 and gains 4.28 percent; another child changes the predictor to TAGE-SC-L and loses 0.17 percent, and its own child enlarges the reorder buffer and gains 0.90 percent. The two reorder-buffer edits are drawn thick as twin paths. A box on the right aligns the two parents, subtracts the gains, and concludes that the predictor limits the window's return. Both sides connect to design memory below.}
\label{fig:overview}
\end{figure}
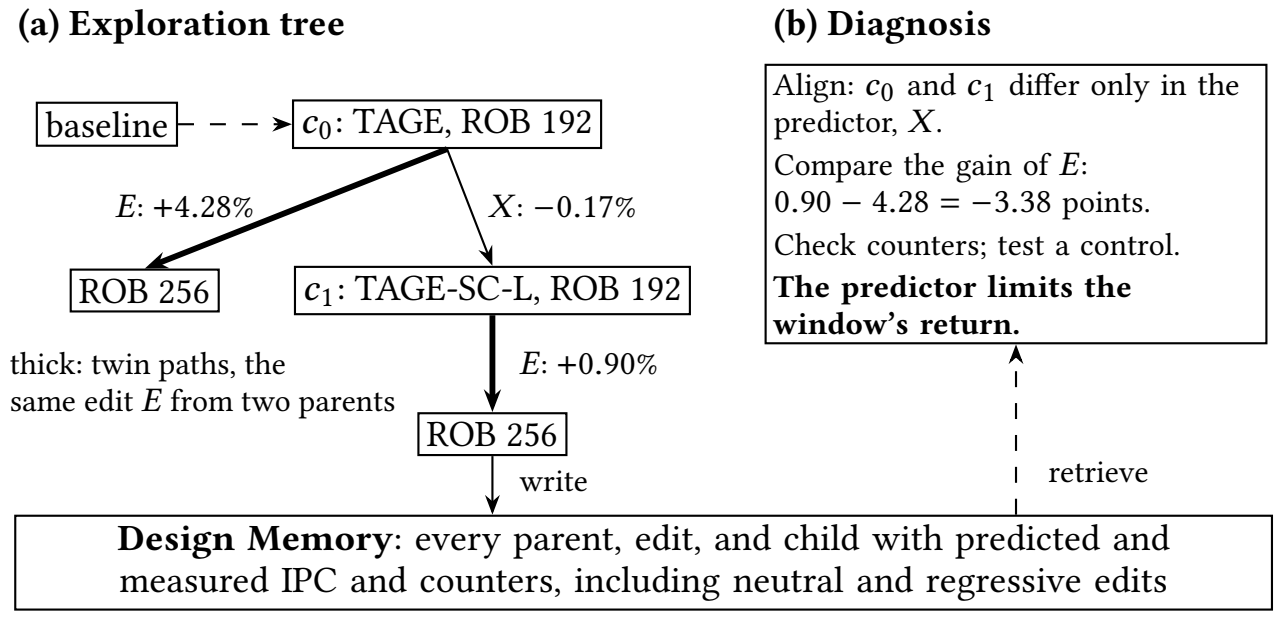

DELTA uses the response to an edit, together with its starting configuration, as the unit of architectural analysis. An edit changes one selected parameter or policy; its parent is the configuration before that change and its child is the configuration after it. Successive edits form a path, also called a trajectory or lineage, and paths that share a parent form a tree (Figure~\ref{fig:overview}). Exploration proposes and measures edits; diagnosis compares their responses across contexts. Both operate on Design Memory, the stored collection of these linked records. A pretrained language model with fixed weights proposes each edit and predicts its IPC; gem5 measurements determine the outcome.

For diagnosis, we use gem5 event counts and derived metrics, including IPC; we refer to these measurements collectively as counters. A pipeline slot is one opportunity to process an operation in a cycle. Top-Down assigns slots to useful retirement, discarded speculative work, insufficient frontend supply, or backend waiting on core or memory causes. These categories name where progress stops; a design change tests what removes that stop.

The prefetch policies, which predict future memory addresses, include the Irregular Stream Buffer (ISB)~\cite{isb_micro13}, Delta Correlating Prediction Tables (DCPT)~\cite{grannaes2011dcpt}, and Slim Access Map Pattern Matching (SlimAMPM). MPP-TAGE combines perceptron prediction with TAGE. Record, Guide, Plan, and Full below identify DELTA variants.

\subsection{Records and exploration}
\label{sec:record}\label{sec:explore}
A record contains the parent configuration $c$, edit $E$, child $E(c)$, predicted and measured IPC, complete available counter vector, trajectory parent, and rationale. A counter vector lists the measurements for one execution; the configuration vector lists the machine settings. The measurement contract is the set of conditions held equal: simulator version, CPU, clock, workloads, inputs, SimPoints, warm-up, aggregation, and repetitions. SimPoints are representative program intervals whose weights approximate the full workload; warm-up executes preceding work to establish simulation state before measurement. An edit signature identifies its old and new values and timing assumptions: increasing cache capacity at fixed latency is a different comparison from increasing capacity with an added hit cycle. At fixed clock, $g=100(\mathrm{IPC}(E(c))/\mathrm{IPC}(c)-1)$ classifies an edit as helpful (H), neutral (N), or regressive (R), with a $\pm0.5\%$ neutral band.

Each exploration iteration retrieves related recorded paths, outcome counts per parameter, parent slot profiles in feedback variants, and previous predictions with their measurements. The language model returns a parent, one edit, predicted child IPC, and rationale. Planning selects among three proposals rather than requesting one. The checker verifies bounds, fixed settings, and duplicates, returning invalid proposals for repair before simulation. Every measured child enters memory, including neutral and regressive outcomes. The model uses these records without task-specific supervised training, backpropagation, or fine-tuning.

\subsection{Twin comparison and root-cause elimination}
\label{sec:alignment}\label{sec:walkthrough}
Diagnosis asks which resource limits the benefit of changing another resource. Three levels distinguish what a run measures, what an edit changes, and how that response depends on the parent context. Let $z_k(c,w)$ be counter $k$ measured on configuration $c$ for workload $w$, and let $E(c)$ apply edit $E$ to that configuration. Define
\[
\delta_k(E\mid c,w)=z_k(E(c),w)-z_k(c,w).
\]
The delta is the child's measurement minus its parent's. Twin paths apply the same $E$ to two parents, $c_0$ and $c_1$, that differ only in one context field $X$, such as the branch predictor used before a ROB increase (Figure~\ref{fig:overview}). Their difference is
\[
T_k(E,X,w)=\delta_k(E\mid c_1,w)-\delta_k(E\mid c_0,w).
\]
For IPC, the same comparison uses four measured values $Y_{ab}$, with $a=0/1$ denoting before/after the edit and $b=0/1$ denoting the first/second context. The four measurements give $I=(Y_{11}-Y_{01})-(Y_{10}-Y_{00})=T_Y$. Thus $I$ subtracts the edit's gain in the first context from its gain in the second. A positive $I$ means the second context increases that gain; zero means equal gains at the reported precision. These four configurations form a comparison rectangle, with one axis for the edit and one for context. Other configuration differences remain explicit; a recorded contrast becomes an isolated interaction when the four configurations satisfy the same measurement contract. Shared corners are not independent repetitions.

Algorithm~\ref{alg:explore} gives the general elimination procedure. A recurring counter response identifies an association to examine. A helpful--regressive reversal pauses that exploration branch and initiates diagnosis. Each candidate mechanism must explain both paths through a resource dependency and necessary counter conditions. A contradictory condition eliminates that candidate; an unresolved condition retains it. Unchanged event counts alone leave service time and overlap as separate possibilities.

The order matters. First establish equal completed work, so that longer execution is not caused by a larger task. Next ask what each proposed cause requires: exhausted capacity requires waiting for that capacity, while insufficient request supply can leave capacity idle. Compare the corresponding counts and waiting times on both paths. Then change the resource that separates the remaining explanations and check whether both the predicted counter response and the performance response occur. A sign reversal selects the comparison; this sequence supplies the explanation.

\begin{algorithm}[t]
\caption{General root-cause elimination. Twin counter differences include IPC; necessary conditions and measured controls determine which resource limits another.}
\label{alg:explore}
\small
\begin{algorithmic}[1]
\Require Records, architectural question, counter meanings
\State Retrieve repeated edits; check work and measurement contracts.
\State Preserve raw counters, units, scope, and each denominator.
\State Compare responses; pause a branch at a speedup reversal.
\State Align twin parents; record every differing configuration field.
\If{the selected context field is not isolated}
  \State Return the association and the required matched comparison.
\EndIf
\State Compute every available $T_k$, including IPC.
\State Form candidate mechanisms that can explain both paths.
\ForAll{necessary conditions that distinguish candidates}
  \State Read counter levels and deltas; eliminate only conflicts.
  \State Retain unresolved candidates and the observation needed.
\EndFor
\ForAll{unused measured controls that separate candidates}
  \State Check fixed settings; compare counter and performance responses.
  \State Eliminate inconsistent candidates; follow remaining stalls.
\EndFor
\State Return the supported conditional dependency, or next control.
\end{algorithmic}
\end{algorithm}

Counters retain raw values and the denominator used for normalization: events per completed work, stalls per work and per cycle, and service rates in their native units. IPC belongs to this vector. Top-Down fractions partition issue slots, so relieving one stall can increase another fraction while execution becomes faster. DELTA therefore checks a fraction's numerator and denominator before interpreting it as increased pressure. A change in a slot fraction is reported in percentage points; an IPC gain is a relative percentage. Missing counters remain unknown.

A separating control connects the explanation to a design decision. If changing resource $R$ reduces the predicted waiting in $S$ and improves performance, its response supports ``$R$ limits $S$'' in that context. If waiting falls without a gain, diagnosis follows the remaining stalls. The model interprets counters and proposes comparisons; deterministic checks align records, and the architect implements controls. Only measured controls update the evidence.

\section{Experimental Methodology}
\label{sec:methodology}
The evaluation separates architectural diagnosis, control studies, and reuse of measurements for search and prediction. Each study retains its own workload, baseline, and comparison conditions; their results are not pooled.

\begin{table}[t]
\centering\footnotesize\setlength{\tabcolsep}{2.5pt}\setlength{\arrayrulewidth}{.35pt}

\begin{tabularx}{\linewidth}{|l|X|}
\hline
Component & Configuration\\\hline
Simulator & gem5 24.0.0.1\\\hline
Core & x86 O3, 8-wide, 2\,GHz\\\hline
Window and queues & ROB 192; Issue Queue (IQ) 64; load/store queues 128 each\\\hline
Registers & 256 integer and 256 floating-point\\\hline
Caches & 32\,KB, 8-way L1I and L1D; 2\,MB L2\\\hline
Workloads & SPEC CPU2017: gcc, mcf, xalancbmk, deepsjeng\\\hline
Sampling & Ten weighted SimPoints per workload; 40 slices\\\hline
Baseline aggregate IPC & 0.96350\\\hline
\end{tabularx}
\caption{Nominal gem5 baseline and 40 weighted slices across four SPEC CPU2017 workloads for the 100 edits.}
\label{tab:machine}
\end{table}

\paragraph{Exploration histories.}
The four SPEC CPU2017 workloads in Table~\ref{tab:machine} use ten weighted SimPoints each~\cite{simpoint_asplos02} in gem5~\cite{gem5_v20}. \stalllens derives Top-Down fractions~\cite{yasin_ispass14}; against hardware counters on 22 workloads, it matches the dominant category in 17. DELTA-Record requests one proposal without bottleneck feedback; Guide adds feedback; Plan selects among three proposals; Full combines both. Each contributes 25 edits, for 100 edges. A suffix identifies an edit in its variant; ``base'' is the baseline.

Edits cover predictors, prefetchers, cache capacity and ways, windows, queues, registers, and frontend width. Nominal limits are 128\,KB per L1 and 16\,MB L2; extended probes reach 256--512\,KB L1s, ROB 512, IQ 128, and Load Queue (LQ) 192. Capacity edits fix hit latency.

\paragraph{Comparison conditions.}
\label{sec:screen}\label{app:screen}
\noindent A retained edge passes the checks used for comparisons across contexts. A held edge remains in memory with its measured outcome but requires a matched comparison before assigning that outcome to a context interaction.
S1 checks that identical configurations reproduce their IPC across variants. S2 checks parent stability across different children. S3 checks that the recorded edit changes the simulated machine. These rules retain 57 edges (24 H/14 N/19 R) for comparisons across contexts and hold 43 as recorded candidate comparisons: S1 covers 25, S2 covers 10, and S3 covers 9, with Guide:18 shared by S1 and S2. S4 pools slots using slot counts as weights; the ratio of IPC to the Retiring slot fraction (expressed from 0 to 1) spans 3.02--3.98.

\paragraph{Search and prediction.}
Search baselines use ten seeds and four benchmarks with the same checker. Random samples valid edits, the genetic algorithm (GA) changes one configuration field (a gene) and selects by IPC, and Top-Down Greedy maps the best parent's dominant stall to a corresponding structure. For search quality, we report the IPC gain after ten iterations. For search cost, we count completed simulations until IPC reaches 1.44 ($1.65\times$ the reference IPC 0.8744). Initial information and search policies remain specific to each method. The broader 25-workload study uses 10\,M warm-up and 100\,M simulation instructions, from harmonic mean IPC 0.4828.

The controlled width study uses gem5 26.0.0.0. Fetch, decode, rename, dispatch, issue, writeback, and commit bandwidths change together from four to eight; paired settings cross ROB 64/192, L1D 16/64\,KiB, and LocalBP/TAGE, with 256\,KiB L2 and 50\,ns memory. Eight microbenchmarks, short programs chosen to exercise particular processor behavior, cover dependent and independent additions, predictable and random branches, and 24\,KiB/1\,MiB pointer traversal with one/eight streams. The four-wide source and eight-wide target machines each supply 64 measurements. Target designs (64, 16, LocalBP), then (192, 64, TAGE), provide 0/8/16 support measurements; each tuple gives ROB entries, L1D KiB, and branch predictor. Support measurements are target examples supplied before prediction; the remaining six designs provide the same 48 test queries. DELTA uses fixed GPT-5 weights, architecture descriptions, history, counters, and prediction feedback. Baselines use configuration and workload features: networks with one hidden layer of 8/32 units (NN8/NN32, five seeds), and a 300-tree random forest (RF). They keep their fitted parameters fixed or retrain on source plus support labels; constant training features are suppressed at test time. Test labels and counters are reserved for scoring. Reported mean absolute percentage error (MAPE) is averaged over  two DELTA repetitions and five NN seeds.

\paragraph{Control studies.}
Recovery tests use the eight-wide core, ROB 192, and L1D 16\,KiB; memory pairs use ROB 64/192, 4/8/16 streams, and two/four MSHRs. Checksums and committed instruction counts match within pairs. Zero added control delay isolates the response of recovery and admission. Section~\ref{sec:transformer} states the transformer controller, work interval, and energy boundary.

\section{Architectural Interactions and Controls}
\label{sec:results}
We ask how design context changes the value of an edit, which counters distinguish the resulting limits, and which controls remove them. Each result reads the tree of Figure~\ref{fig:overview}, first across recorded edits, then through matched controls and repairs.

\subsection{Which edits help, and where do gains saturate?}
\label{sec:taxonomy}\label{sec:regularities}
Added capacity rarely lowers IPC but stops paying at a size that differs per structure; replacing a predictor or prefetcher gives both the largest gains and most losses.

\noindent Each edit is tagged by what it changes: storage (how much state a structure holds), concurrency (how much work can remain in flight), prediction (the hardware's choice of future control flow or memory requests), and allocation (where entries can be placed). At the $\pm0.5\%$ band, the 100 outcomes are 36 H/30 N/34 R; they are 39/25/36 at $\pm0.25\%$ and 25/44/31 at $\pm1\%$, so about one edit in three lowers IPC at every band.

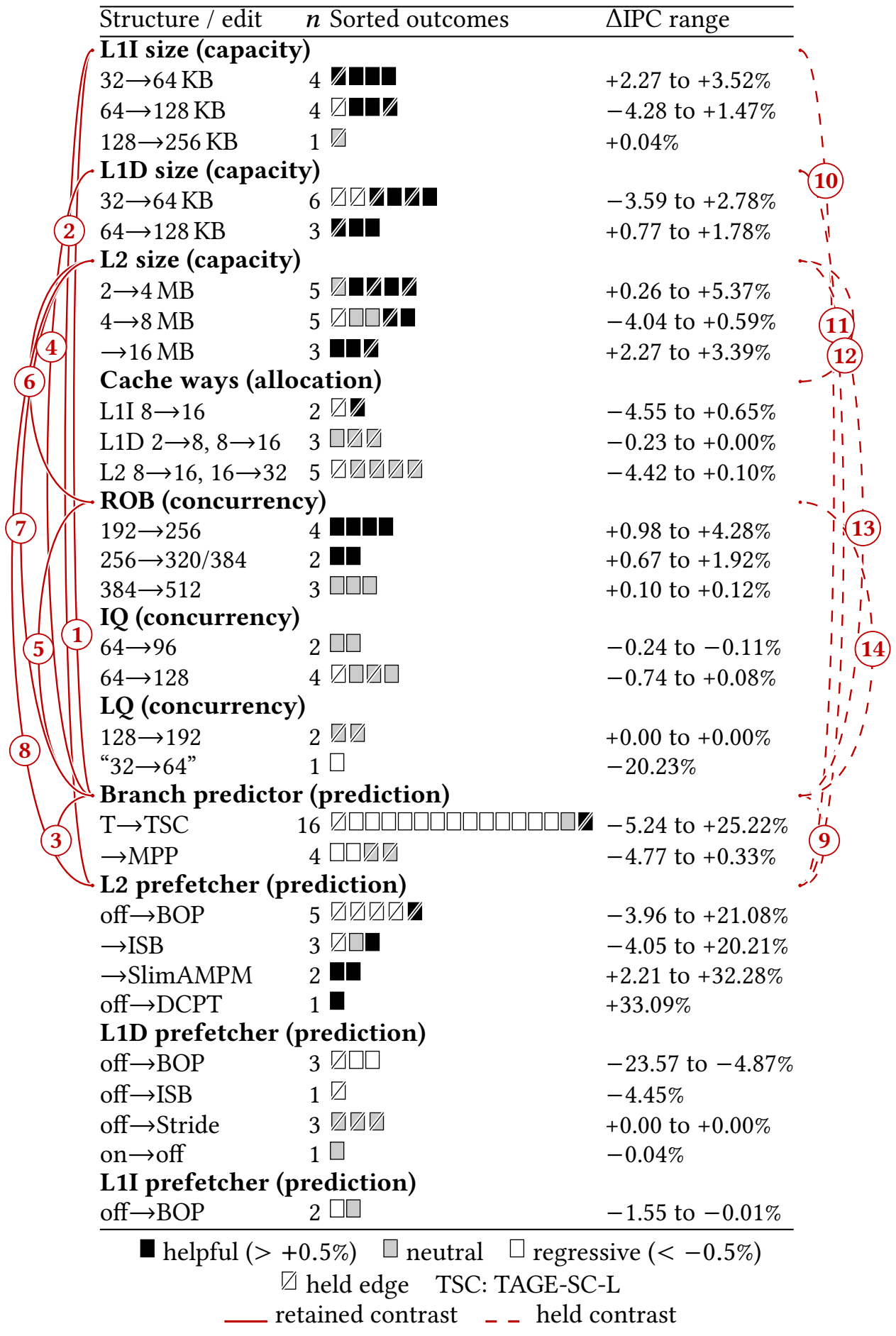
\begin{figure}[t]
\centering\footnotesize
\newcommand{\edgecell}[2]{\tikz[baseline=-.3mm]{
\ifnum#1=1\def\shade{black}\else\ifnum#1=2\def\shade{black!20}\else\def\shade{white}\fi\fi
\draw[fill=\shade,line width=.25pt] (0,0) rectangle (1.25mm,1.6mm);
\ifnum#2=1\draw[white,line width=1pt] (0,0)--(1.25mm,1.6mm);
\draw[black,line width=.25pt] (0,0)--(1.25mm,1.6mm);\fi}\hspace{.18mm}}
\setlength{\tabcolsep}{1.2pt}\renewcommand{\arraystretch}{0.94}
\begin{tabular}{@{}lrll@{}}
\hline
Structure / edit & $n$ & Sorted outcomes & $\Delta$IPC range\\\hline
\multicolumn{4}{@{}l}{\tikz[remember picture,baseline]{\coordinate (ed-l1i) at (0,.6ex);}\textbf{L1I size (capacity)}}\\
32$\to$64\,KB & 4 & \edgecell{1}{1}\edgecell{1}{0}\edgecell{1}{0}\edgecell{1}{0} & +2.27 to +3.52\%\\
64$\to$128\,KB & 4 & \edgecell{3}{1}\edgecell{1}{0}\edgecell{1}{0}\edgecell{1}{1} & $-$4.28 to +1.47\%\\
128$\to$256\,KB & 1 & \edgecell{2}{1} & +0.04\%\\
\multicolumn{4}{@{}l}{\tikz[remember picture,baseline]{\coordinate (ed-l1d) at (0,.6ex);}\textbf{L1D size (capacity)}}\\
32$\to$64\,KB & 6 & \edgecell{3}{1}\edgecell{3}{1}\edgecell{1}{1}\edgecell{1}{0}\edgecell{1}{1}\edgecell{1}{0} & $-$3.59 to +2.78\%\\
64$\to$128\,KB & 3 & \edgecell{1}{1}\edgecell{1}{0}\edgecell{1}{0} & +0.77 to +1.78\%\\
\multicolumn{4}{@{}l}{\tikz[remember picture,baseline]{\coordinate (ed-l2) at (0,.6ex);}\textbf{L2 size (capacity)}}\\
2$\to$4\,MB & 5 & \edgecell{2}{1}\edgecell{1}{0}\edgecell{1}{1}\edgecell{1}{0}\edgecell{1}{1} & +0.26 to +5.37\%\\
4$\to$8\,MB & 5 & \edgecell{3}{1}\edgecell{2}{0}\edgecell{2}{0}\edgecell{1}{1}\edgecell{1}{0} & $-$4.04 to +0.59\%\\
$\to$16\,MB & 3 & \edgecell{1}{0}\edgecell{1}{0}\edgecell{1}{1} & +2.27 to +3.39\%\\
\multicolumn{4}{@{}l}{\tikz[remember picture,baseline]{\coordinate (ed-ways) at (0,.6ex);}\textbf{Cache ways (allocation)}}\\
L1I 8$\to$16 & 2 & \edgecell{3}{1}\edgecell{1}{1} & $-$4.55 to +0.65\%\\
L1D 2$\to$8, 8$\to$16 & 3 & \edgecell{2}{0}\edgecell{2}{1}\edgecell{2}{1} & $-$0.23 to +0.00\%\\
L2 8$\to$16, 16$\to$32 & 5 & \edgecell{3}{1}\edgecell{2}{1}\edgecell{2}{1}\edgecell{2}{1}\edgecell{2}{1} & $-$4.42 to +0.10\%\\
\multicolumn{4}{@{}l}{\tikz[remember picture,baseline]{\coordinate (ed-rob) at (0,.6ex);}\textbf{ROB (concurrency)}}\\
192$\to$256 & 4 & \edgecell{1}{0}\edgecell{1}{0}\edgecell{1}{0}\edgecell{1}{0} & +0.98 to +4.28\%\\
256$\to$320/384 & 2 & \edgecell{1}{0}\edgecell{1}{0} & +0.67 to +1.92\%\\
384$\to$512 & 3 & \edgecell{2}{0}\edgecell{2}{0}\edgecell{2}{0} & +0.10 to +0.12\%\\
\multicolumn{4}{@{}l}{\tikz[remember picture,baseline]{\coordinate (ed-iq) at (0,.6ex);}\textbf{IQ (concurrency)}}\\
64$\to$96 & 2 & \edgecell{2}{0}\edgecell{2}{0} & $-$0.24 to $-$0.11\%\\
64$\to$128 & 4 & \edgecell{3}{1}\edgecell{2}{0}\edgecell{2}{1}\edgecell{2}{0} & $-$0.74 to +0.08\%\\
\multicolumn{4}{@{}l}{\tikz[remember picture,baseline]{\coordinate (ed-lq) at (0,.6ex);}\textbf{LQ (concurrency)}}\\
128$\to$192 & 2 & \edgecell{2}{1}\edgecell{2}{1} & +0.00 to +0.00\%\\
``32$\to$64'' & 1 & \edgecell{3}{0} & $-$20.23\%\\
\multicolumn{4}{@{}l}{\tikz[remember picture,baseline]{\coordinate (ed-bp) at (0,.6ex);}\textbf{Branch predictor (prediction)}}\\
T$\to$TSC & 16 & \edgecell{3}{1}\edgecell{3}{0}\edgecell{3}{0}\edgecell{3}{0}\edgecell{3}{0}\edgecell{3}{0}\edgecell{3}{0}\edgecell{3}{0}\edgecell{3}{0}\edgecell{3}{0}\edgecell{3}{0}\edgecell{3}{0}\edgecell{3}{0}\edgecell{3}{0}\edgecell{2}{0}\edgecell{1}{1} & $-$5.24 to +25.22\%\\
$\to$MPP & 4 & \edgecell{3}{0}\edgecell{3}{0}\edgecell{2}{1}\edgecell{2}{1} & $-$4.77 to +0.33\%\\
\multicolumn{4}{@{}l}{\tikz[remember picture,baseline]{\coordinate (ed-l2pf) at (0,.6ex);}\textbf{L2 prefetcher (prediction)}}\\
off$\to$BOP & 5 & \edgecell{3}{1}\edgecell{3}{1}\edgecell{3}{1}\edgecell{3}{1}\edgecell{1}{1} & $-$3.96 to +21.08\%\\
$\to$ISB & 3 & \edgecell{3}{1}\edgecell{2}{0}\edgecell{1}{0} & $-$4.05 to +20.21\%\\
$\to$SlimAMPM & 2 & \edgecell{1}{0}\edgecell{1}{0} & +2.21 to +32.28\%\\
off$\to$DCPT & 1 & \edgecell{1}{0} & +33.09\%\\
\multicolumn{4}{@{}l}{\tikz[remember picture,baseline]{\coordinate (ed-l1dpf) at (0,.6ex);}\textbf{L1D prefetcher (prediction)}}\\
off$\to$BOP & 3 & \edgecell{3}{1}\edgecell{3}{0}\edgecell{3}{0} & $-$23.57 to $-$4.87\%\\
off$\to$ISB & 1 & \edgecell{3}{1} & $-$4.45\%\\
off$\to$Stride & 3 & \edgecell{2}{1}\edgecell{2}{1}\edgecell{2}{1} & +0.00 to +0.00\%\\
on$\to$off & 1 & \edgecell{2}{0} & $-$0.04\%\\
\multicolumn{4}{@{}l}{\tikz[remember picture,baseline]{\coordinate (ed-l1ipf) at (0,.6ex);}\textbf{L1I prefetcher (prediction)}}\\
off$\to$BOP & 2 & \edgecell{3}{0}\edgecell{2}{0} & $-$1.55 to $-$0.01\%\\
\hline
\end{tabular}
\begin{tikzpicture}[remember picture,overlay]
\tikzset{ic/.style={draw=red!75!black,line width=.45pt,preaction={draw=white,line width=1.1pt}}}
\draw[ic] ($(ed-bp)+(-0.7mm,0)$) .. controls ($(ed-bp)+(-3.0mm,0)$) and ($(ed-l1i)+(-3.0mm,0)$) .. ($(ed-l1i)+(-0.7mm,0)$);
\fill[red!75!black] ($(ed-bp)+(-0.7mm,0)$) circle (.45pt);
\fill[red!75!black] ($(ed-l1i)+(-0.7mm,0)$) circle (.45pt);
\draw[ic] ($(ed-l1i)+(-0.7mm,0)$) .. controls ($(ed-l1i)+(-4.1mm,0)$) and ($(ed-l2pf)+(-4.1mm,0)$) .. ($(ed-l2pf)+(-0.7mm,0)$);
\fill[red!75!black] ($(ed-l1i)+(-0.7mm,0)$) circle (.45pt);
\fill[red!75!black] ($(ed-l2pf)+(-0.7mm,0)$) circle (.45pt);
\draw[ic] ($(ed-bp)+(-0.7mm,0)$) .. controls ($(ed-bp)+(-5.2mm,0)$) and ($(ed-l2pf)+(-5.2mm,0)$) .. ($(ed-l2pf)+(-0.7mm,0)$);
\fill[red!75!black] ($(ed-bp)+(-0.7mm,0)$) circle (.45pt);
\fill[red!75!black] ($(ed-l2pf)+(-0.7mm,0)$) circle (.45pt);
\draw[ic] ($(ed-bp)+(-0.7mm,0)$) .. controls ($(ed-bp)+(-6.300000000000001mm,0)$) and ($(ed-l1d)+(-6.300000000000001mm,0)$) .. ($(ed-l1d)+(-0.7mm,0)$);
\fill[red!75!black] ($(ed-bp)+(-0.7mm,0)$) circle (.45pt);
\fill[red!75!black] ($(ed-l1d)+(-0.7mm,0)$) circle (.45pt);
\draw[ic] ($(ed-bp)+(-0.7mm,0)$) .. controls ($(ed-bp)+(-7.4mm,0)$) and ($(ed-rob)+(-7.4mm,0)$) .. ($(ed-rob)+(-0.7mm,0)$);
\fill[red!75!black] ($(ed-bp)+(-0.7mm,0)$) circle (.45pt);
\fill[red!75!black] ($(ed-rob)+(-0.7mm,0)$) circle (.45pt);
\draw[ic] ($(ed-l2)+(-0.7mm,0)$) .. controls ($(ed-l2)+(-8.5mm,0)$) and ($(ed-rob)+(-8.5mm,0)$) .. ($(ed-rob)+(-0.7mm,0)$);
\fill[red!75!black] ($(ed-l2)+(-0.7mm,0)$) circle (.45pt);
\fill[red!75!black] ($(ed-rob)+(-0.7mm,0)$) circle (.45pt);
\draw[ic] ($(ed-bp)+(-0.7mm,0)$) .. controls ($(ed-bp)+(-9.600000000000001mm,0)$) and ($(ed-l2)+(-9.600000000000001mm,0)$) .. ($(ed-l2)+(-0.7mm,0)$);
\fill[red!75!black] ($(ed-bp)+(-0.7mm,0)$) circle (.45pt);
\fill[red!75!black] ($(ed-l2)+(-0.7mm,0)$) circle (.45pt);
\draw[ic] ($(ed-l2pf)+(-0.7mm,0)$) .. controls ($(ed-l2pf)+(-10.700000000000001mm,0)$) and ($(ed-l2)+(-10.700000000000001mm,0)$) .. ($(ed-l2)+(-0.7mm,0)$);
\fill[red!75!black] ($(ed-l2pf)+(-0.7mm,0)$) circle (.45pt);
\fill[red!75!black] ($(ed-l2)+(-0.7mm,0)$) circle (.45pt);
\draw[ic,dashed] ($(ed-bp)+(65.3mm,0)$) .. controls ($(ed-bp)+(68.3mm,0)$) and ($(ed-l2pf)+(68.3mm,0)$) .. ($(ed-l2pf)+(65.3mm,0)$);
\fill[red!75!black] ($(ed-bp)+(65.3mm,0)$) circle (.45pt);
\fill[red!75!black] ($(ed-l2pf)+(65.3mm,0)$) circle (.45pt);
\draw[ic,dashed] ($(ed-l1i)+(65.3mm,0)$) .. controls ($(ed-l1i)+(69.5mm,0)$) and ($(ed-l2pf)+(69.5mm,0)$) .. ($(ed-l2pf)+(65.3mm,0)$);
\fill[red!75!black] ($(ed-l1i)+(65.3mm,0)$) circle (.45pt);
\fill[red!75!black] ($(ed-l2pf)+(65.3mm,0)$) circle (.45pt);
\draw[ic,dashed] ($(ed-l1d)+(65.3mm,0)$) .. controls ($(ed-l1d)+(70.7mm,0)$) and ($(ed-l2pf)+(70.7mm,0)$) .. ($(ed-l2pf)+(65.3mm,0)$);
\fill[red!75!black] ($(ed-l1d)+(65.3mm,0)$) circle (.45pt);
\fill[red!75!black] ($(ed-l2pf)+(65.3mm,0)$) circle (.45pt);
\draw[ic,dashed] ($(ed-ways)+(65.3mm,0)$) .. controls ($(ed-ways)+(71.89999999999999mm,0)$) and ($(ed-l2)+(71.89999999999999mm,0)$) .. ($(ed-l2)+(65.3mm,0)$);
\fill[red!75!black] ($(ed-ways)+(65.3mm,0)$) circle (.45pt);
\fill[red!75!black] ($(ed-l2)+(65.3mm,0)$) circle (.45pt);
\draw[ic,dashed] ($(ed-bp)+(65.3mm,0)$) .. controls ($(ed-bp)+(73.1mm,0)$) and ($(ed-l2)+(73.1mm,0)$) .. ($(ed-l2)+(65.3mm,0)$);
\fill[red!75!black] ($(ed-bp)+(65.3mm,0)$) circle (.45pt);
\fill[red!75!black] ($(ed-l2)+(65.3mm,0)$) circle (.45pt);
\draw[ic,dashed] ($(ed-bp)+(65.3mm,0)$) .. controls ($(ed-bp)+(74.3mm,0)$) and ($(ed-rob)+(74.3mm,0)$) .. ($(ed-rob)+(65.3mm,0)$);
\fill[red!75!black] ($(ed-bp)+(65.3mm,0)$) circle (.45pt);
\fill[red!75!black] ($(ed-rob)+(65.3mm,0)$) circle (.45pt);
\path ($(ed-bp)+(-0.7mm,0)$) .. controls ($(ed-bp)+(-3.0mm,0)$) and ($(ed-l1i)+(-3.0mm,0)$) .. ($(ed-l1i)+(-0.7mm,0)$) node[pos=0.3,circle,fill=white,draw=red!75!black,line width=.35pt,text=red!75!black,inner sep=.5pt,minimum size=2.8mm,font=\fontsize{6.5}{7}\selectfont\bfseries] {1};
\path ($(ed-l1i)+(-0.7mm,0)$) .. controls ($(ed-l1i)+(-4.1mm,0)$) and ($(ed-l2pf)+(-4.1mm,0)$) .. ($(ed-l2pf)+(-0.7mm,0)$) node[pos=0.3,circle,fill=white,draw=red!75!black,line width=.35pt,text=red!75!black,inner sep=.5pt,minimum size=2.8mm,font=\fontsize{6.5}{7}\selectfont\bfseries] {2};
\path ($(ed-bp)+(-0.7mm,0)$) .. controls ($(ed-bp)+(-5.2mm,0)$) and ($(ed-l2pf)+(-5.2mm,0)$) .. ($(ed-l2pf)+(-0.7mm,0)$) node[pos=0.5,circle,fill=white,draw=red!75!black,line width=.35pt,text=red!75!black,inner sep=.5pt,minimum size=2.8mm,font=\fontsize{6.5}{7}\selectfont\bfseries] {3};
\path ($(ed-bp)+(-0.7mm,0)$) .. controls ($(ed-bp)+(-6.300000000000001mm,0)$) and ($(ed-l1d)+(-6.300000000000001mm,0)$) .. ($(ed-l1d)+(-0.7mm,0)$) node[pos=0.65,circle,fill=white,draw=red!75!black,line width=.35pt,text=red!75!black,inner sep=.5pt,minimum size=2.8mm,font=\fontsize{6.5}{7}\selectfont\bfseries] {4};
\path ($(ed-bp)+(-0.7mm,0)$) .. controls ($(ed-bp)+(-7.4mm,0)$) and ($(ed-rob)+(-7.4mm,0)$) .. ($(ed-rob)+(-0.7mm,0)$) node[pos=0.5,circle,fill=white,draw=red!75!black,line width=.35pt,text=red!75!black,inner sep=.5pt,minimum size=2.8mm,font=\fontsize{6.5}{7}\selectfont\bfseries] {5};
\path ($(ed-l2)+(-0.7mm,0)$) .. controls ($(ed-l2)+(-8.5mm,0)$) and ($(ed-rob)+(-8.5mm,0)$) .. ($(ed-rob)+(-0.7mm,0)$) node[pos=0.5,circle,fill=white,draw=red!75!black,line width=.35pt,text=red!75!black,inner sep=.5pt,minimum size=2.8mm,font=\fontsize{6.5}{7}\selectfont\bfseries] {6};
\path ($(ed-bp)+(-0.7mm,0)$) .. controls ($(ed-bp)+(-9.600000000000001mm,0)$) and ($(ed-l2)+(-9.600000000000001mm,0)$) .. ($(ed-l2)+(-0.7mm,0)$) node[pos=0.5,circle,fill=white,draw=red!75!black,line width=.35pt,text=red!75!black,inner sep=.5pt,minimum size=2.8mm,font=\fontsize{6.5}{7}\selectfont\bfseries] {7};
\path ($(ed-l2pf)+(-0.7mm,0)$) .. controls ($(ed-l2pf)+(-10.700000000000001mm,0)$) and ($(ed-l2)+(-10.700000000000001mm,0)$) .. ($(ed-l2)+(-0.7mm,0)$) node[pos=0.3,circle,fill=white,draw=red!75!black,line width=.35pt,text=red!75!black,inner sep=.5pt,minimum size=2.8mm,font=\fontsize{6.5}{7}\selectfont\bfseries] {8};
\path ($(ed-bp)+(65.3mm,0)$) .. controls ($(ed-bp)+(68.3mm,0)$) and ($(ed-l2pf)+(68.3mm,0)$) .. ($(ed-l2pf)+(65.3mm,0)$) node[pos=0.5,circle,fill=white,draw=red!75!black,line width=.35pt,text=red!75!black,inner sep=.5pt,minimum size=2.8mm,font=\fontsize{6.5}{7}\selectfont\bfseries] {9};
\path ($(ed-l1i)+(65.3mm,0)$) .. controls ($(ed-l1i)+(69.5mm,0)$) and ($(ed-l2pf)+(69.5mm,0)$) .. ($(ed-l2pf)+(65.3mm,0)$) node[pos=0.25,circle,fill=white,draw=red!75!black,line width=.35pt,text=red!75!black,inner sep=.5pt,minimum size=2.8mm,font=\fontsize{6.5}{7}\selectfont\bfseries] {10};
\path ($(ed-l1d)+(65.3mm,0)$) .. controls ($(ed-l1d)+(70.7mm,0)$) and ($(ed-l2pf)+(70.7mm,0)$) .. ($(ed-l2pf)+(65.3mm,0)$) node[pos=0.3,circle,fill=white,draw=red!75!black,line width=.35pt,text=red!75!black,inner sep=.5pt,minimum size=2.8mm,font=\fontsize{6.5}{7}\selectfont\bfseries] {11};
\path ($(ed-ways)+(65.3mm,0)$) .. controls ($(ed-ways)+(71.89999999999999mm,0)$) and ($(ed-l2)+(71.89999999999999mm,0)$) .. ($(ed-l2)+(65.3mm,0)$) node[pos=0.3,circle,fill=white,draw=red!75!black,line width=.35pt,text=red!75!black,inner sep=.5pt,minimum size=2.8mm,font=\fontsize{6.5}{7}\selectfont\bfseries] {12};
\path ($(ed-bp)+(65.3mm,0)$) .. controls ($(ed-bp)+(73.1mm,0)$) and ($(ed-l2)+(73.1mm,0)$) .. ($(ed-l2)+(65.3mm,0)$) node[pos=0.5,circle,fill=white,draw=red!75!black,line width=.35pt,text=red!75!black,inner sep=.5pt,minimum size=2.8mm,font=\fontsize{6.5}{7}\selectfont\bfseries] {13};
\path ($(ed-bp)+(65.3mm,0)$) .. controls ($(ed-bp)+(74.3mm,0)$) and ($(ed-rob)+(74.3mm,0)$) .. ($(ed-rob)+(65.3mm,0)$) node[pos=0.5,circle,fill=white,draw=red!75!black,line width=.35pt,text=red!75!black,inner sep=.5pt,minimum size=2.8mm,font=\fontsize{6.5}{7}\selectfont\bfseries] {14};
\end{tikzpicture}
\par\smallskip
\edgecell{1}{0} helpful ($>+0.5\%$)\quad
\edgecell{2}{0} neutral\quad \edgecell{3}{0} regressive ($<-0.5\%$)\\
\edgecell{3}{1} held edge\quad TSC: TAGE-SC-L\\
\tikz{\draw[red!75!black,line width=.6pt](0,0)--(4mm,0);} retained contrast\quad \tikz{\draw[red!75!black,dashed,line width=.6pt](0,0)--(4mm,0);} held contrast
\caption{All 100 edits; curves connect resource pairs compared in Section~\ref{sec:interactions}, including the worked L1I--ISB interaction (curve 2). Dashed curves include held records; diagonals mark the 43 held edges.}
\Description{The figure groups 100 edits by the changed structure, displaying outcome marks, counts, IPC ranges, and comparison scope. Solid red curves in the left margin connect component rows with contrasts across four configurations on retained records. Dashed red curves in the right margin identify candidate comparisons that require matched configuration and measurement conditions before causal interpretation. Curves are numbered 1 through 14 and identify component pairs. Section~\ref{sec:interactions} explains the interactions, including the four IPC measurements and calculation for the L1I and ISB example on curve 2. Crossings route separate curves.}
\label{fig:edittree}
\end{figure}

Those losses, called regressions, concentrate in one family. Prediction and speculation edits account for 18 of the 19 retained regressions (4 H/4 N/18 R), against 14 H/2 N/0 R for storage and 6 H/7 N/1 R for concurrency; allocation contributes one neutral edge. 
A larger structure at the same latency retains the smaller one’s state; a new predictor or prefetcher replaces all of the old policy’s decisions, including correct ones.
Across all 100 records, storage gives 23/4/4, concurrency 6/10/2, prediction/speculation 6/9/26, and allocation 1/7/2 (Figure~\ref{fig:edittree}). Concurrency edits also carry a storage tag, yielding 49 capacity tags.

\paragraph{Which records can be compared across contexts?}
Three checks decide. Identical configurations must reproduce their IPC (S1): Guide's baseline TAGE-SC-L IPC is 1.20647 against 0.93064 in Record and Plan, and the 64\,KB L1I values are 1.23381 and 0.95632. A parent must be stable across its children (S2): Record:5's $+5.37\%$ L2 step shifts Frontend Bound, Bad Speculation, and Memory Bound by $+4.47$, $-3.99$, and $-2.46$ points, and its children inherit the inverse shifts whatever their edit. The edit must change the simulated machine (S3), which covers four unchanged associativity settings, three stride edits with zero effect, and two LQ edits to an inactive queue. Thus recorded sign changes identify comparisons to resolve, while retained twin comparisons measure context dependence.

Evidence labels distinguish observed IPC responses, associated counter changes, mechanisms supported by intervention, and effects replicated in further contexts. The classification has 15 explicit mechanism labels; eight edges pair IPC with counter evidence.

\paragraph{Where capacity gains stop.}
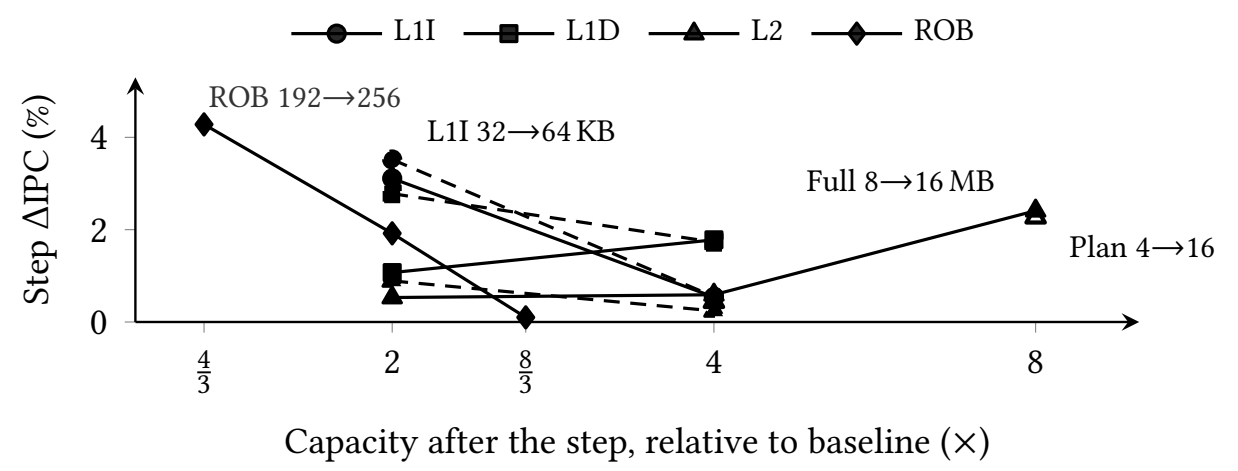
\begin{figure}[t]
\centering
\begin{tikzpicture}\begin{axis}[
width=.97\linewidth,height=32mm,xmode=log,log basis x=2,
xmin=1.15,xmax=10,ymin=0,ymax=5.3,
xtick={1.3333,2,2.6667,4,8},xticklabels={{\scriptsize$\frac{4}{3}$},2,{\scriptsize$\frac{8}{3}$},4,8},
xlabel={Capacity after the step, relative to baseline ($\times$)},
ylabel={Step $\Delta$IPC (\%)},label style={font=\footnotesize},
tick label style={font=\footnotesize},
grid=none,
legend style={font=\scriptsize,draw=none,fill=white,fill opacity=.85,text opacity=1,at={(.5,1.03)},anchor=south,legend columns=4,/tikz/every even column/.append style={column sep=3pt}},
clip=false]
\addplot[black,semithick,mark=*,mark size=1.6pt] coordinates {(2,3.11) (4,0.54)};
\addplot[black,semithick,mark=square*,mark size=1.5pt] coordinates {(2,1.07) (4,1.78)};
\addplot[black,semithick,mark=triangle*,mark size=2pt] coordinates {(2,0.53) (4,0.59) (8,2.41)};
\addplot[black,semithick,mark=diamond*,mark size=2pt] coordinates {(1.3333,4.28) (2,1.92) (2.6667,0.10)};
\legend{L1I,L1D,L2,ROB}
\addplot[black,semithick,densely dashed,mark=*,mark size=1.6pt,forget plot] coordinates {(2,3.52) (4,0.55)};
\addplot[black,semithick,densely dashed,mark=square*,mark size=1.5pt,forget plot] coordinates {(2,2.78) (4,1.73)};
\addplot[black,semithick,densely dashed,mark=triangle*,mark size=2pt,forget plot] coordinates {(2,0.89) (4,0.24)};
\addplot[ink,only marks,mark=triangle,mark size=2pt,thick,forget plot] coordinates {(4,0.44) (8,2.27)};
\node[font=\scriptsize,text=ink,anchor=south east] at (axis cs:7.6,2.47) {Full 8$\rightarrow$16\,MB};
\node[font=\scriptsize,text=ink,anchor=north west] at (axis cs:8.3,2.25) {Plan 4$\rightarrow$16};
\node[font=\scriptsize,text=accent,anchor=south west] at (axis cs:2.08,3.55) {L1I 32$\rightarrow$64\,KB};
\node[font=\scriptsize,text=slate,anchor=south west] at (axis cs:1.3,4.28) {ROB 192$\rightarrow$256};
\node[font=\scriptsize,text=slate,anchor=north] at (axis cs:2.6667,0.0) {};
\end{axis}\end{tikzpicture}
\caption{IPC relative to the parent at the reached capacity: Full (solid), Plan (dashed), and Plan's shared DCPT parent (hollow).}
\Description{Line plot of marginal IPC change versus capacity multiple for L1I, L1D, L2, and ROB steps along the Full and Plan trajectories. L1I and ROB gains fall sharply after the first step, L1D gains stay between one and three percent, and L2 gains rise to 2.41 percent at 16 MB.}
\label{fig:scaling}
\end{figure}

Each structure stops paying at its own size. L1I saturates first: 32 to 64\,KB gains 3.11--3.52\% across three retained contexts (2.27\% in Guide's held edge), and the next doubling adds only 0.54--0.55\%. L1D returns persist: Full and Plan gain 1.78\% and 1.73\% on the second doubling, compared with 1.08\% and 2.78\% on the first. L2 pays late: Full's steps from 2 to 16\,MB gain 0.53\%, 0.59\%, and 2.41\%; at Plan's DCPT parent, 4 to 8\,MB gains 0.44\% while 4 to 16\,MB gains 2.27\%. The capacity return therefore depends on the cache level and working set (Figure~\ref{fig:scaling}).

The window also saturates: ROB expansion helps through 384 entries ($+0.66$ to $+4.28\%$), then flattens at 512 ($+0.10$--$0.12\%$); IQ expansion is neutral in four retained contexts. The sole retained concurrency regression, Plan:23 ($-20.23\%$), places the LQ below its 128-entry baseline and raises Core Bound from 0.229 to 0.499. No retained associativity edit leaves the neutral band.

\begin{figure}[t]
\centering
\begin{tikzpicture}
\begin{axis}[width=\linewidth,height=28mm,ybar,bar width=9pt,
 ymin=-12,ymax=6.6,xmin=0.45,xmax=4.55,
 xtick={1,2,3,4},
 xticklabels={{Full:4\\ROB\,256 | TSC},{Full:1\\L1D\,64\,KB | BOP},{Record:18\\ROB\,320 | TSC},{Plan:16\\ROB\,256 | BOP}},
 ylabel={$\Delta$IPC from parent (\%)},ylabel style={font=\footnotesize},
 tick label style={font=\scriptsize},x tick label style={align=center},
 axis line style={black,line width=.45pt},grid=none,
 legend style={font=\scriptsize,draw=none,fill=white,fill opacity=.85,text opacity=1,at={(.5,.98)},anchor=north,legend columns=2,/tikz/every even column/.append style={column sep=4pt}},
 legend image code/.code={\draw[#1] (0cm,-0.08cm) rectangle (0.22cm,0.12cm);},
 nodes near coords,every node near coord/.append style={font=\scriptsize,/pgf/number format/fixed,/pgf/number format/fixed zerofill,/pgf/number format/precision=2,/pgf/number format/showpos=true},
 clip=false]
\addplot[fill=black,draw=black] coordinates {(1,4.28) (2,1.08) (3,0.66) (4,1.02)};
\addplot[fill=white,draw=black,pattern=north east lines,pattern color=black] coordinates {(1,-0.17) (2,-6.71) (3,-4.23) (4,-4.86)};
\draw[slate] (axis cs:0.45,0)--(axis cs:4.55,0);
\end{axis}\end{tikzpicture}
\caption{Paired edits from four shared parents: capacity/window (solid) versus predictor/L1D BOP (hatched). TSC: TAGE-SC-L.}
\Description{Grouped bars for four shared parents. Capacity or window edits gain 4.28, 1.08, 0.66, and 1.02 percent. The competing predictor or L1D prefetcher edits change IPC by minus 0.17, minus 6.71, minus 4.23, and minus 4.86 percent.}
\label{fig:sharedparent}
\end{figure}
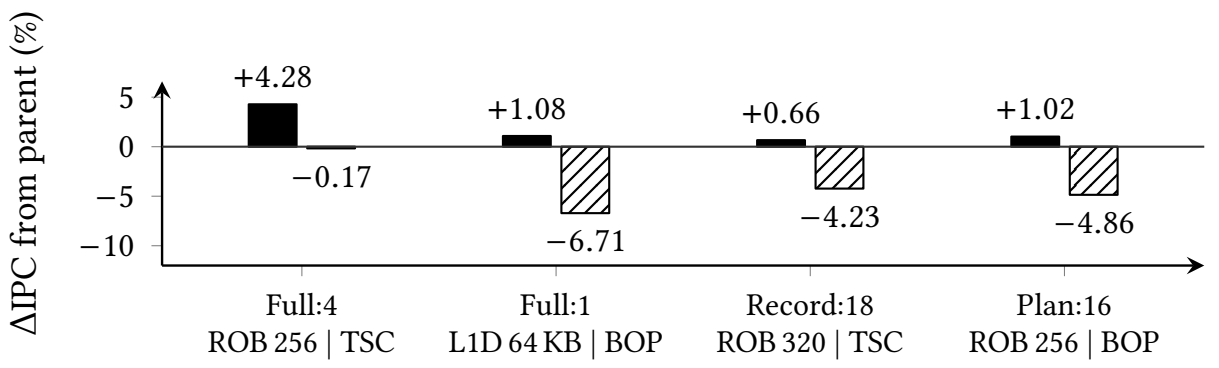

Children of a shared parent start from identical stalls, so their difference comes from the edit alone (Figure~\ref{fig:sharedparent}). Buffering helps in all four pairs while predictor or L1D BOP replacement is neutral or regressive: at Full:4, ROB growth gains 4.28\% versus $-0.17\%$ for TAGE-SC-L; at Full:1, L1D growth gains 1.08\% versus $-6.71\%$ for BOP. Identical parent slot profiles therefore admit different outcomes: slot relief alone does not predict IPC.

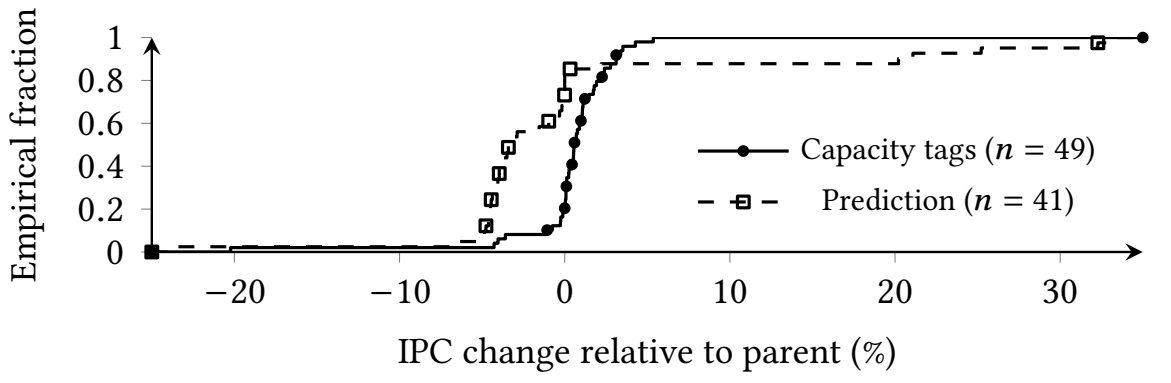
\begin{figure}[t]\centering
\begin{tikzpicture}\begin{axis}[width=.96\linewidth,height=30mm,xmin=-25,xmax=35,ymin=0,ymax=1,
xlabel={IPC change relative to parent (\%)},ylabel={Empirical fraction},
label style={font=\footnotesize},tick label style={font=\footnotesize},
legend style={font=\scriptsize,draw=none,fill=white,fill opacity=.85,text opacity=1,at={(.98,.06)},anchor=south east,legend columns=1},
grid=none]
\addplot[black,solid,const plot,semithick,mark=*,mark size=.8pt,mark repeat=5] coordinates {(-25.000000,0.000000) (-20.231559,0.020408) (-4.275270,0.040816) (-4.037667,0.061224) (-3.589479,0.081633) (-1.063905,0.102041) (-0.745480,0.122449) (-0.258577,0.142857) (-0.239404,0.163265) (-0.112820,0.183673) (0.000000,0.204082) (0.000000,0.224490) (0.043855,0.244898) (0.075765,0.265306) (0.075765,0.285714) (0.101746,0.306122) (0.113625,0.326531) (0.121181,0.346939) (0.241028,0.367347) (0.263131,0.387755) (0.444593,0.408163) (0.525339,0.428571) (0.528446,0.448980) (0.539721,0.469388) (0.553074,0.489796) (0.592824,0.510204) (0.664911,0.530612) (0.705677,0.551020) (0.766299,0.571429) (0.885733,0.591837) (0.982875,0.612245) (1.015791,0.632653) (1.063739,0.653061) (1.075171,0.673469) (1.108551,0.693878) (1.219799,0.714286) (1.470086,0.734694) (1.733754,0.755102) (1.783777,0.775510) (1.924660,0.795918) (2.264867,0.816327) (2.266115,0.836735) (2.407288,0.857143) (2.778412,0.877551) (3.114686,0.897959) (3.114686,0.918367) (3.393790,0.938776) (3.521262,0.959184) (4.279025,0.979592) (5.366278,1.000000) (35.000000,1.000000)};
\addplot[black,dashed,const plot,semithick,mark=square,mark size=1.1pt,mark repeat=5,mark options={solid,fill=white}] coordinates {(-25.000000,0.000000) (-23.568023,0.024390) (-6.712778,0.048780) (-5.240782,0.073171) (-4.864924,0.097561) (-4.768238,0.121951) (-4.610296,0.146341) (-4.586366,0.170732) (-4.585770,0.195122) (-4.492657,0.219512) (-4.448952,0.243902) (-4.384546,0.268293) (-4.268913,0.292683) (-4.228396,0.317073) (-4.046424,0.341463) (-3.962801,0.365854) (-3.961213,0.390244) (-3.912509,0.414634) (-3.743294,0.439024) (-3.495006,0.463415) (-3.410483,0.487805) (-3.410483,0.512195) (-3.403168,0.536585) (-2.882940,0.560976) (-1.550525,0.585366) (-0.968760,0.609756) (-0.624710,0.634146) (-0.309211,0.658537) (-0.170137,0.682927) (-0.037349,0.707317) (-0.008689,0.731707) (-0.006227,0.756098) (0.000000,0.780488) (0.000000,0.804878) (0.000000,0.829268) (0.325781,0.853659) (2.211234,0.878049) (20.211170,0.902439) (21.083550,0.926829) (25.217436,0.951220) (32.279494,0.975610) (33.087895,1.000000) (35.000000,1.000000)};
\legend{Capacity tags ($n=49$),Prediction ($n=41$)}
\end{axis}\end{tikzpicture}
\caption{IPC responses for capacity and prediction tags in the 100-edge store, including held edges. Prediction and speculation edits have a larger tail of IPC losses, while edits tagged as capacity have a positive median.}
\Description{Empirical cumulative distributions compare outcomes tagged capacity and prediction across recorded edges.}\label{fig:cdf}\end{figure}

Across all records, edits tagged as capacity have a $+0.59\%$ median and six losses among 49 cases; edits tagged as prediction have a $-3.41\%$ median, 26 losses among 41 cases, and all five largest gains ($20$--$33\%$; Figure~\ref{fig:cdf}). TAGE-SC-L is regressive in 13 of 14 retained contexts and neutral in one; MPP-TAGE is regressive in both retained contexts. By contrast, the first L2 prefetcher supplies the largest jump outside Guide: ISB 20.21\%, SlimAMPM 32.28\%, and DCPT 33.09\%. L1D prefetching helps in none of eight edges, and L1I BOP changes IPC by $-1.55\%$ and $-0.01\%$. The same broad family thus contains both the largest gains and frequent losses.

\subsection{Which resources must be chosen together?}
\label{sec:interactions}\label{app:interactions}
\noindent Instruction supply changes the benefit of prefetching, and branch prediction changes the return from a larger instruction window. We call this coupling: changing one resource changes the measured gain from another. The numbered curves in Figure~\ref{fig:edittree} identify the resource pairs: curves 1--8 use retained records, while curves 9--14 identify candidate comparisons under S1 or S2.

\paragraph{Instruction supply enables prefetching.}
For L1I capacity and ISB (curve 2), the four IPC measurements give a worked interaction. With 32\,KB L1I, enabling ISB changes IPC from 0.96350 to 0.96344; with 64\,KB L1I, it changes IPC from 0.99351 to 1.19431. The corresponding increments are $-0.00006$ and $+0.20080$ IPC. Their difference is $I=0.20080-(-0.00006)=+0.20086$ IPC, following Section~\ref{sec:alignment}. Thus ISB is neutral at 32\,KB L1I ($-0.01\%$) but gains 20.21\% at 64\,KB. L1I growth alone gains 3.11\%, whereas the combination gains 23.96\%. Frontend Bound falls 2.77 points after ISB in the larger cache, while Bad Speculation rises 5.04 points. Faster data service becomes valuable once the frontend supplies enough work; the larger speculation share accompanies higher throughput.

\paragraph{Branch prediction changes the return from a window.}
With 64\,KB L1 caches and SlimAMPM, Full's ROB gains under TAGE versus TAGE-SC-L are 4.28\% versus 0.90\% for 192 to 256, 6.29\% versus 1.56\% for 192 to 384, and 1.92\% versus 0.65\% for 256 to 384 (curve 5). Predictor replacement changes IPC by $-0.17\%$, $-3.40\%$, and $-4.61\%$ at ROB 192, 256, and 384. A nearly neutral predictor change in a small window suppresses much of the return from a larger window. Window capacity and the prediction context supplying it must therefore be evaluated together.

\paragraph{Memory context separates strong and weak coupling.}
ISB reduces the predictor penalty from 3.74\% to 0.97\%, and its Bad Speculation increase from 2.70 to 0.14 points. DCPT versus SlimAMPM instead preserves similar penalties, 4.59\% and 4.38\% (curve 3). L1D gains rise from 1.08\% to 1.54\% under predictor replacement in Full, but fall from 2.78\% to 2.69\% in Plan (curve 4). The rest of the configuration can reverse this coupling.

Small interactions also guide design. ROB 384 to 512 gains 0.10\%, 0.12\%, and 0.11\% at L2 sizes 2, 4, and 16\,MB (curve 6). L2 4 to 16\,MB gains 3.01\% and 2.91\% under the two predictors (curve 7). At Plan's 4\,MB parent, DCPT gains 33.09\%, while doubling capacity gains 0.24\% without DCPT and 0.44\% with it (curve 8). Timely service matters more than storage there.

\paragraph{Recorded contrasts select the next comparison.}
Under Guide's S1 condition, BOP changes from a 21.08\% gain to a 3.96\% loss with predictor replacement (curve 9). Curves 10--14 record further contrasts in L1I, L1D, L2, and ROB gains. These comparisons preserve their measured responses while directing the next comparison to a common root, stable parent, and matched executed prefetch settings.

\subsection{Can stall fractions identify the limiting resource?}
\label{sec:slots}
A stall fraction alone does not determine the performance response: similar IPC changes can have different counter signatures, and aggregate gains can hide workload differences. L1I growth lowers Frontend Bound by 2.20 points in three retained contexts; 11 TAGE-SC-L edges raise Bad Speculation by 2.42--3.01 points; two MPP-TAGE edges agree within 0.13 points across slots (Figure~\ref{fig:signatures}). Across those 11 TAGE-SC-L edges, Core Bound drops 0.6--1.5 points: IPC losses are 3.4--3.7\% at four parents without L2 prefetching, versus 4.2--4.6\% at six of seven with it. From Full's 16\,MB L2 parent, TAGE-SC-L loses 4.59\% with Bad Speculation up 2.98 points, while MPP-TAGE loses 4.77\% with Frontend Bound up 2.25 points. Similar throughput losses therefore point to different next controls.

\begin{figure}[t]
\centering
\definecolor{regress}{HTML}{707070}
\begin{tikzpicture}[x=1mm,y=1mm,font=\scriptsize]
\node[font=\tiny\bfseries,text=black] at (26.65,1.60) {Frontend};
\node[font=\tiny\bfseries,text=black] at (34.95,1.60) {Bad Spec.};
\node[font=\tiny\bfseries,text=black] at (43.25,1.60) {Core};
\node[font=\tiny\bfseries,text=black] at (51.55,1.60) {Memory};
\node[font=\tiny\bfseries,text=black] at (59.85,1.60) {Retiring};
\node[font=\tiny\bfseries,text=black] at (70.75,1.60) {$\Delta$IPC (\%)};
\node[font=\tiny\bfseries,text=black] at (79.25,1.60) {$n$};
\node[anchor=east,text=black,inner sep=0.5pt] at (21.90,-1.52) {L1I 32$\rightarrow$64\,KB};
\node[text=black] at (26.65,-1.52) {$-$2.2};
\node[text=black] at (34.95,-1.52) {+0.7};
\node[text=black] at (43.25,-1.52) {+0.6};
\node[text=black] at (51.55,-1.52) {0.0};
\node[text=black] at (59.85,-1.52) {+1.0};
\node[text=black] at (70.75,-1.52) {+3.1 to +3.5};
\node[text=black] at (79.25,-1.52) {3};
\node[anchor=east,text=black,inner sep=0.5pt] at (21.90,-4.57) {L1I 64$\rightarrow$128\,KB};
\node[text=black] at (26.65,-4.57) {$-$0.4};
\node[text=black] at (34.95,-4.57) {+0.2};
\node[text=black] at (43.25,-4.57) {+0.1};
\node[text=black] at (51.55,-4.57) {0.0};
\node[text=black] at (59.85,-4.57) {+0.1};
\node[text=black] at (70.75,-4.57) {+0.5};
\node[text=black] at (79.25,-4.57) {2};
\node[anchor=east,text=black,inner sep=0.5pt] at (21.90,-7.62) {L1D enlargement};
\node[text=black] at (26.65,-7.62) {+0.2};
\node[text=black] at (34.95,-7.62) {$-$0.2};
\node[text=black] at (43.25,-7.62) {$-$0.3};
\node[text=black] at (51.55,-7.62) {$-$0.2};
\node[text=black] at (59.85,-7.62) {+0.4};
\node[text=black] at (70.75,-7.62) {+1.1 to +2.8};
\node[text=black] at (79.25,-7.62) {4};
\node[anchor=east,text=black,inner sep=0.5pt] at (21.90,-10.67) {L2 enlargement};
\node[text=black] at (26.65,-10.67) {0.0};
\node[text=black] at (34.95,-10.67) {0.0};
\node[text=black] at (43.25,-10.67) {0.0};
\node[text=black] at (51.55,-10.67) {0.0};
\node[text=black] at (59.85,-10.67) {+0.1};
\node[text=black] at (70.75,-10.67) {+0.2 to +2.4};
\node[text=black] at (79.25,-10.67) {7};
\node[anchor=east,text=black,inner sep=0.5pt] at (21.90,-14.62) {ROB to 384};
\node[text=black] at (26.65,-14.62) {+0.3};
\node[text=black] at (34.95,-14.62) {+0.2};
\node[text=black] at (43.25,-14.62) {$-$0.9};
\node[text=black] at (51.55,-14.62) {+0.1};
\node[text=black] at (59.85,-14.62) {+0.3};
\node[text=black] at (70.75,-14.62) {+0.7 to +4.3};
\node[text=black] at (79.25,-14.62) {6};
\node[anchor=east,text=black,inner sep=0.5pt] at (21.90,-17.67) {ROB 384$\rightarrow$512};
\node[text=black] at (26.65,-17.67) {0.0};
\node[text=black] at (34.95,-17.67) {0.0};
\node[text=black] at (43.25,-17.67) {0.0};
\node[text=black] at (51.55,-17.67) {0.0};
\node[text=black] at (59.85,-17.67) {+0.1};
\node[text=black] at (70.75,-17.67) {+0.1};
\node[text=black] at (79.25,-17.67) {3};
\node[anchor=east,text=black,inner sep=0.5pt] at (21.90,-20.72) {IQ enlargement};
\node[text=black] at (26.65,-20.72) {+0.3};
\node[text=black] at (34.95,-20.72) {+1.2};
\node[text=black] at (43.25,-20.72) {$-$0.9};
\node[text=black] at (51.55,-20.72) {$-$0.5};
\node[text=black] at (59.85,-20.72) {$-$0.1};
\node[text=black] at (70.75,-20.72) {$-$0.3 to +0.1};
\node[text=black] at (79.25,-20.72) {4};
\node[anchor=east,text=black,inner sep=0.5pt] at (21.90,-24.67) {First L2 prefetcher};
\node[text=black] at (26.65,-24.67) {$-$2.1};
\node[text=black] at (34.95,-24.67) {+5.0};
\node[text=black] at (43.25,-24.67) {$-$3.6};
\node[text=black] at (51.55,-24.67) {$-$2.3};
\node[text=black] at (59.85,-24.67) {+3.1};
\node[text=black] at (70.75,-24.67) {+20 to +33};
\node[text=black] at (79.25,-24.67) {3};
\node[anchor=east,text=black,inner sep=0.5pt] at (21.90,-27.72) {L1D BOP};
\node[text=black] at (26.65,-27.72) {$-$0.4};
\node[text=black] at (34.95,-27.72) {+0.1};
\node[text=black] at (43.25,-27.72) {+0.6};
\node[text=black] at (51.55,-27.72) {+0.5};
\node[text=black] at (59.85,-27.72) {$-$0.9};
\node[text=black] at (70.75,-27.72) {$-$6.7 to $-$4.9};
\node[text=black] at (79.25,-27.72) {2};
\node[anchor=east,text=black,inner sep=0.5pt] at (21.90,-30.77) {TAGE-SC-L};
\node[text=black] at (26.65,-30.77) {$-$0.3};
\node[text=black] at (34.95,-30.77) {+2.7};
\node[text=black] at (43.25,-30.77) {$-$0.8};
\node[text=black] at (51.55,-30.77) {$-$0.2};
\node[text=black] at (59.85,-30.77) {$-$1.2};
\node[text=black] at (70.75,-30.77) {$-$4.6 to $-$0.2};
\node[text=black] at (79.25,-30.77) {14};
\node[anchor=east,text=black,inner sep=0.5pt] at (21.90,-33.82) {MPP-TAGE};
\node[text=black] at (26.65,-33.82) {+2.2};
\node[text=black] at (34.95,-33.82) {+1.4};
\node[text=black] at (43.25,-33.82) {$-$1.9};
\node[text=black] at (51.55,-33.82) {$-$0.1};
\node[text=black] at (59.85,-33.82) {$-$1.6};
\node[text=black] at (70.75,-33.82) {$-$4.8 to $-$4.3};
\node[text=black] at (79.25,-33.82) {2};
\draw[black,line width=.35pt] (1.50,-12.65)--(81.00,-12.65);
\draw[black,line width=.35pt] (1.50,-22.70)--(81.00,-22.70);
\draw[black,line width=.4pt] (1.5,3.2) rectangle (81,-35.6);
\draw[black,line width=.35pt] (1.5,-.05)--(81,-.05);
\draw[black,line width=.35pt] (22.4,3.2)--(22.4,-35.6);
\draw[black,line width=.35pt] (64.2,3.2)--(64.2,-35.6);
\draw[black,line width=.35pt] (77,3.2)--(77,-35.6);
\end{tikzpicture}
\caption{Median slot changes (signed percentage points), IPC ranges relative to the parent, and edge counts.}
\Description{Ruled matrix of median changes in slot fractions in Frontend Bound, Bad Speculation, Core Bound, Memory Bound, and Retiring for eleven retained edit types, with the range of IPC change relative to the parent and the edge count of each type.}
\label{fig:signatures}
\end{figure}

ROB expansion lowers Core Bound by 0.76--1.40 points in five of six retained contexts. At Full:4, ROB growth lowers Bad Speculation by 4.45 points and gains 4.28\%, whereas TAGE-SC-L lowers that share by 1.80 points with $-0.17\%$ IPC. Across nine retained cache gains, Memory Bound changes by at most 0.20 points. Initial L2 prefetchers lower Core Bound by 2.61--4.54 points; SlimAMPM gains 32.28\% while Bad Speculation rises from 14.00\% to 19.84\%. Slot fractions locate changes in the distribution of execution, while IPC determines the net response.

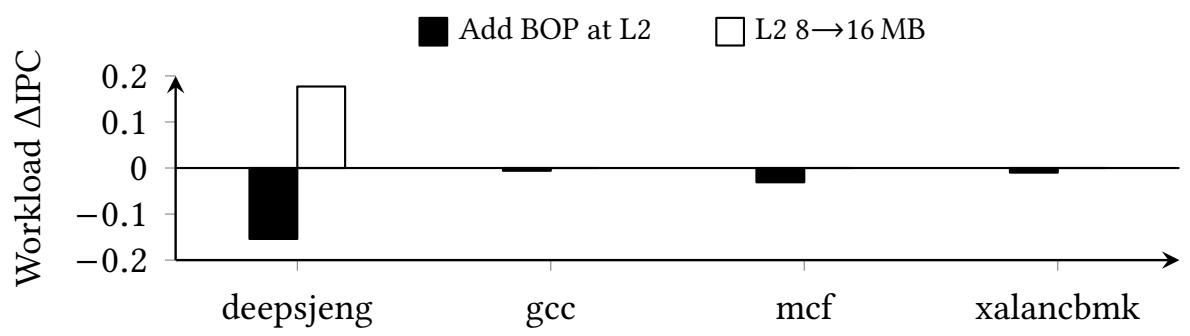
\begin{figure}[t]
\centering
\begin{tikzpicture}\begin{axis}[
width=.97\linewidth,height=28mm,ybar=0pt,bar width=9pt,ymin=-.20,ymax=.20,
ytick={-.2,-.1,0,.1,.2},
symbolic x coords={deepsjeng,gcc,mcf,xalancbmk},xtick=data,enlarge x limits=.16,
ylabel={Workload $\Delta$IPC},label style={font=\footnotesize},tick label style={font=\footnotesize},
xticklabel style={text height=1.5ex,text depth=.4ex},
axis line style={black,line width=.45pt},grid=none,
legend image code/.code={\draw[#1] (0pt,-2.6pt) rectangle (5.2pt,2.6pt);},
legend style={font=\scriptsize,draw=none,fill=none,
  at={(.5,1.04)},anchor=south,legend columns=2,
  /tikz/every even column/.append style={column sep=10pt}}]
\addplot[fill=black,draw=black] coordinates {(deepsjeng,-.154)(gcc,-.006)(mcf,-.031)(xalancbmk,-.010)};
\addplot[fill=white,draw=black] coordinates {(deepsjeng,.177)(gcc,0)(mcf,0)(xalancbmk,0)};
\legend{Add BOP at L2,L2 8$\rightarrow$16\,MB}
\draw[black,line width=.4pt]
  ({rel axis cs:0,0}|-{axis cs:deepsjeng,0}) -- ({rel axis cs:1,0}|-{axis cs:deepsjeng,0});
\end{axis}\end{tikzpicture}
\caption{Workload responses within Guide: rounded IPC changes from each edit's own parent.}
\Description{Paired black and open bars give workload IPC changes for adding L2 BOP and growing L2 capacity from 8 to 16 MB. The horizontal reference line marks zero change.}
\label{fig:workloadcontrast}
\end{figure}

Workload separation identifies effects hidden by aggregation (Figure~\ref{fig:workloadcontrast}). Within Guide, deepsjeng supplies 76--99\% of four BOP losses of 2.88--3.96\%; mcf loses 0.019--0.032\,IPC. At Guide:5, gcc/xalancbmk gain 0.012/0.006\,IPC while deepsjeng/mcf lose 0.141/0.019. Growing L2 from 8 to 16\,MB instead raises deepsjeng from 1.356 to 1.533\,IPC without changing the others, although its initial Memory Bound share is only 0.069, versus 0.227 Frontend Bound and 0.230 Bad Speculation. A small initial fraction does not determine the return from the corresponding resource.

\subsection{Does wider execution require faster recovery?}
\label{sec:recovery}
A wider pipeline admits more instructions behind each unresolved branch, so a misprediction leaves more wrong-path work for the ROB to discard. Microbenchmarks with random branches make that cost visible. At ROB 192, widening from four to eight changes IPC by $-8.78\%$ with LocalBP and $+6.04\%$ with TAGE; at ROB 64, the changes are $-2.07\%$ and $+10.73\%$. With ROB 192, L1D 16\,KiB, and LocalBP, mispredictions fall by 1.46\% while cycles rise by 9.63\%: the wider core mispredicts no more often but pays more for each misprediction.

\begin{table}[t]
\centering\footnotesize\setlength{\tabcolsep}{2.5pt}\setlength{\arrayrulewidth}{.35pt}

\setlength{\tabcolsep}{2.5pt}
\begin{tabularx}{\linewidth}{|l|X|r|}
\hline
Predictor & Intervention & Speedup\\\hline
LocalBP & Conditional admission & 5.66\%\\\hline
LocalBP & Branch admission & 12.86\%\\\hline
TAGE & Squash width 32 & 21.03\%\\\hline
LocalBP & Squash width 32 & 29.38\%\\\hline
\end{tabularx}
\caption{Random branches, ROB 192; speedup is the ratio of baseline to intervention cycles, minus one.}
\label{tab:recovery}
\end{table}

Raising ROB squash width from 8 to 32 gains 29.38\% with LocalBP and 21.03\% with TAGE (Table~\ref{tab:recovery}); width 64 gains 37.95\% and 25.67\%. LocalBP squash cycles fall by 84.08\% while mispredictions rise by 1.49\%, so faster removal of wrong-path work, rather than better prediction, explains this response. These microbenchmarks test recovery apart from the SPEC predictor--ROB interaction.

The alternative is to admit less work. Branch admission reduces rename width from eight to two when at least two ROB branches remain unresolved, gaining 12.86\%. Conditional admission enables that gate only when squashing occupied at least 30\% of the preceding 256 active rename cycles. It gains 5.66\% and leaves predictable branches and independent additions unchanged. 
With identical loop bytes placed at offset 0 or 32 of a 64-byte block (seeds 23 and 37), squash width gains 29.50–31.01\% when the loop is aligned but only 1.13–1.15\% at offset 32, where branch addresses also change.
Conditional admission gains 5.39--5.61\% when aligned and is inactive at offset 32.

\paragraph{Does request capacity limit contended loads?}
Three contended load cases identify a separate limit: doubling physical MSHRs from two to four gains 99.19--99.87\% with the admission policy fixed; speed doubles with the entries, so requests in flight are the limit. Together, these controls distinguish admission of work, removal of wrong-path work, and memory requests in flight.

\subsection{When should a prefetcher reset its learning history?}
\label{sec:abop}
The L2 Best-Offset prefetcher (BOP)~\cite{bop_hpca16} learns from the requests that reach L2, and L1D capacity decides which requests those are. An offset is the address displacement used to propose another line; BOP scores candidate offsets from recent requests and prefetches with the best one. A sequence of L1D capacity changes gains 6.8\% from 64 to 128\,KB, while 256\,KB is neutral ($-0.30\%$): more L1D hits filter requests before they reach L2, so the learned scores describe a stream that BOP no longer observes. The architect uses this path comparison to add two controls. Usefulness control checks whether requests selected by that learning actually help; retraining clears the old scores when the observed stream changes. Adaptive BOP improves mean IPC by 1.37\% over original BOP across the reported matched configurations.

\begin{figure}[tb]
\centering
\begin{tikzpicture}[font=\footnotesize,
b/.style={draw=black,fill=white,align=center,minimum height=8mm,minimum width=15mm,inner sep=2pt},
n/.style={draw=black,double,double distance=.55pt,fill=white,line width=.45pt,align=center,minimum height=8mm,minimum width=15mm,inner sep=2pt},
a/.style={-{Stealth[length=1.4mm]},draw=slate},fa/.style={-{Stealth[length=1.4mm]},draw=accent,line width=.6pt}]
\node[b] (l) {Offset\\learning};
\node[b,right=5mm of l] (g) {Best score\\gate};
\node[n,right=5mm of g] (u) {Usefulness\\gate};
\node[b,right=5mm of u] (i) {Prefetch\\issue};
\draw[a] (l)--(g); \draw[a] (g)--(u); \draw[a] (u)--(i);
\draw[fa] (i.north) -- ++(0,3.2mm) -| node[pos=.25,above,font=\scriptsize,text=accent]{useful/issued per 256-access epoch} (u.north);
\draw[fa,dashed] (u.south) -- ++(0,-3.2mm) -| node[pos=.25,below,font=\scriptsize,text=accent]{after two low epochs: clear scores, retrain} (l.south);
\end{tikzpicture}
\caption{Adaptive BOP keeps offset learning and its score gate (single borders); usefulness controls issue, retraining (dashed).}
\Description{A row of four blocks: offset learning, best score gate, a new usefulness gate, and prefetch issue. A feedback arrow returns the ratio of useful to issued prefetches per epoch to the usefulness gate, and a dashed arrow from the gate back to offset learning clears scores and retrains after two low epochs.}
\label{fig:abop}
\end{figure}
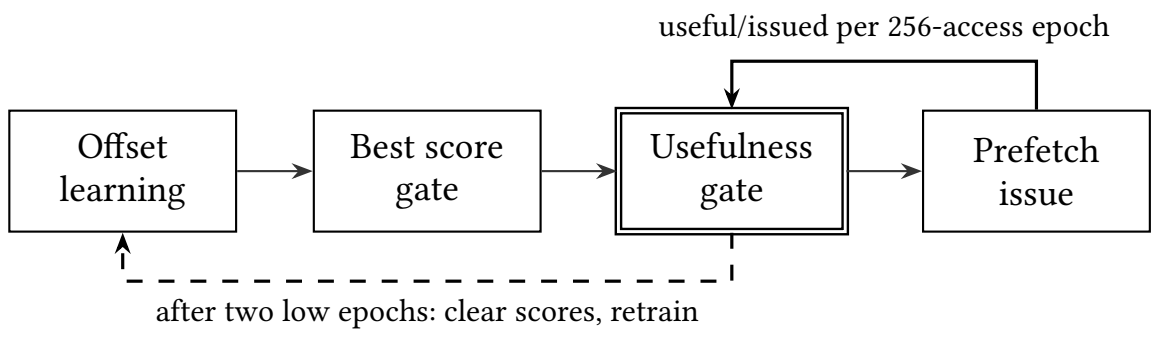

Offset learning is retained: 46 candidate offsets, two 64-entry banks of recent requests, a 60-cycle delay queue, maximum score 31, up to 100 rounds, and bad score threshold 10. Usefulness is the fraction of issued prefetches that are used, measured over each epoch of 256 L2 accesses. Below 0.15, issue probability falls 25 percentage points, to a 25\% floor; above 0.35 it rises 25 points. Two consecutive low epochs clear scores and recent requests, so learning restarts on the current stream (Figure~\ref{fig:abop}). Feedback~\cite{fdp_hpca07} thus governs both issue rate and the lifetime of the learner's history, independently of the offset score.

\subsection{Why can more accurate prefetching slow a transformer?}
\label{sec:transformer}\label{sec:repair}\label{app:transformer}

The workload here is a transformer's forward pass, not the language model that proposes edits. Issuing only the final prefetch target improves accuracy but leaves intervening lines uncovered; filling them shortens execution. The fixed TinyStories 15M transformer in \texttt{llama2.c}~\cite{llama2c} computes next token scores in one FP32 (32-bit floating-point) forward pass through six layers at saved token prefix positions 16/128. Loading, token selection, and output checks are excluded; all 19 runs match native outputs and hashes. Settings are width eight, TAGE, L1I 32\,KiB, L1D 16\,KiB/16 MSHRs, L2 256\,KiB, and 16\,GiB/s bandwidth.

\paragraph{Which controls select request timing?}
Without prefetching, narrowing to four lowers IPC 6.38\% at identical work and 957,875 L2 misses; events with a full IQ rise 644\%. At position 16, L1D 16$\to$64\,KiB, MSHRs 16$\to$4, and squash width 8$\to$32 change cycles less than 0.01\%; admission changes none. These test retention, outstanding requests, and recovery. Bandwidth is 0.896\,GB/s without MSHR exhaustion. Roughly 7.5 million L1D misses and 0.96 million L2 misses instead motivate earlier requests.

DELTA's three rounds with fixed weights, history, and 1,483 counters propose stride degree/distance tests. A stride is the address difference between accesses by one load; degree counts targets, and distance skips strides before the first target. Degree four/distance zero gains 355.1\%/361.4\% at widths four/eight, selecting request timing and coverage with width eight fixed.

\paragraph{Why does accuracy rise while execution slows?}
The architect specifies adaptive lead (distance ahead in strides), issuing only the endpoint. Each load's program counter (PC) identifies its state; a frontier records its farthest proposed address. Lead starts at four: each 64-line window with over eight absent lines increments it up to 16; four windows without absence decrement it toward two. Absent means not resident; ordinary demand hits are excluded from notifications. Fixed lookahead variants lose 15.50\%/8.24\% against stride; adaptive endpoint control loses 38.17\%.

With 1,875 configuration values, simulator binary, work, and output fixed, requests fall 80.47\% (3,683,101$\to$719,479), accuracy rises 0.035653$\to$0.983862, and L1 misses fall 46.50\%, yet cycles rise 61.72\%. An L1D miss either merges with an outstanding request or starts service with a new MSHR. Accuracy divides useful by issued prefetches; coverage divides useful prefetches by useful prefetches plus demand MSHR misses. Accurate requests can therefore leave demands starting service late.

\paragraph{Which counter conditions identify the cause?}
Both runs commit 58,645,013 instructions (ROB writes $-0.1\%$), establishing equal work. Mispredictions change 2,177$\to$2,181 but examined squashed instructions fall 1.2\%, directing attention away from extra recovery work. MSHR blocking falls 185$\to$0 cycles; load queue full events rise 12$\to$43, while full IQ events fall 23.9\%. L2 misses remain similar (958,243$\to$959,714); bandwidth falls 4.07$\to$1.92\,GB/s over longer execution. Coverage falls 0.865$\to$0.744, demand MSHR allocations rise 20,415$\to$243,195 ($11.91\times$), and 250,018 lines are absent, selecting request supply for implementation inspection.

\paragraph{Which state changes restore coverage?}
The frontier advances before queue rejection at a page boundary, suppressing later proposals behind an address never requested. Endpoint issue also skips page prefixes and intervals opened by increasing lead; greater distance fills neither gap. At 40 million instructions, two weight pages have 32 new demand MSHR allocations, no late merges, and 96 ready lines, versus 2/107/19 with stride and 2/0/126 after repair.

DELTA receives counters and implementation descriptions, connects request supply to coverage, and proposes frontier control and gap filling. The architect specifies and implements both: reset progress on page changes and generate every uncovered stride address through the endpoint, stopping at the 4\,KiB boundary or 16 candidates with queue admission unchanged. Page prefixes are initial cache lines, distinct from token prefixes. State adds 71 bits/entry, including a 48-bit frontier (568 bytes/64 entries).

Repair changes the predicted counters: at position 16, candidates crossing pages fall 234,589$\to$0, demand MSHR allocations 243,195$\to$20,871, and coverage rises 0.744$\to$0.978; cycles fall 47,960,555$\to$18,446,082. Requests total 949,680 versus stride's 3,683,101. The joint repair gains 160.0\% over adaptive endpoint control, 60.8\%/55.3\% over stride, and $7.418\times$/$7.092\times$ over no prefetch at positions 16/128, connecting coverage to execution time. The repair carries over to RISC-V: on gem5 cores with the Large/Mega sizes of SonicBOOM~\cite{sonicboom_carrv20}, an open-source out-of-order RISC-V processor, running \texttt{llama2.c} generation, the unchanged controller removes all 170,054/168,481 candidates crossing pages, gaining 4.90\%/4.35\% over endpoint control, 7.84\%/22.80\% over no prefetch, and 1.16\% over stride on Mega.

\paragraph{Does shorter execution reduce modeled energy?}
McPAT~\cite{mcpat}, at 45\,nm, 1\,V, 2\,GHz, 350\,K, estimates core/cache energy reductions from stride of 16.89\%/\allowbreak15.91\% at positions 16/128. Power rises 14.850/\allowbreak14.848$\to$\allowbreak19.842/\allowbreak19.394\,W; dynamic energy falls 2.28\%/\allowbreak2.13\%, so shorter leakage duration dominates. At position 128, reads rise 4.13\% (63,183,488$\to$\allowbreak65,795,072 bytes) and cycles blocked on MSHRs rise 185$\to$4,777. Energy integrates modeled pipeline, predictor, execution unit, and cache power over simulated runtime; DRAM, interconnect, and the new controller's implementation delay and energy remain outside this boundary.

\begin{figure}[t]
\centering
\begin{tikzpicture}\begin{axis}[
width=\linewidth,height=27mm,xtick={0,5,10,15,20,25},ytick={1,1.25,1.5},xmin=0,xmax=25,ymin=.94,ymax=1.55,
xlabel={Design index},ylabel={Best IPC so far},ylabel style={yshift=-3pt},xlabel style={yshift=3pt},
label style={font=\footnotesize},tick label style={font=\footnotesize},
axis line style={black,line width=.45pt},grid=none,
legend style={font=\footnotesize,draw=none,at={(.5,1.03)},anchor=south,legend columns=4,/tikz/every even column/.append style={column sep=4pt}}]
\addplot[black,solid,semithick,mark=*,mark size=.85pt,const plot,mark repeat=4] coordinates {(0,0.9635) (1,0.9935) (2,1.3142) (3,1.3142) (4,1.3142) (5,1.3283) (6,1.3283) (7,1.3852) (8,1.3852) (9,1.4118) (10,1.4118) (11,1.4118) (12,1.4195) (13,1.4448) (14,1.4463) (15,1.4463) (16,1.4524) (17,1.4541) (18,1.4541) (19,1.4610) (20,1.4962) (21,1.4978) (22,1.4978) (23,1.4978) (24,1.4978) (25,1.4978)};
\addplot[black,dashed,semithick,mark=square,mark size=1.1pt,mark options={solid,fill=white},const plot,mark repeat=4] coordinates {(0,0.9635) (1,0.9635) (2,0.9903) (3,0.9903) (4,1.0251) (5,1.0342) (6,1.0399) (7,1.0580) (8,1.0605) (9,1.4080) (10,1.4080) (11,1.4143) (12,1.4143) (13,1.4399) (14,1.4399) (15,1.4399) (16,1.4399) (17,1.4718) (18,1.4718) (19,1.4718) (20,1.4718) (21,1.4718) (22,1.4718) (23,1.4718) (24,1.4718) (25,1.4867)};
\addplot[black,densely dotted,semithick,mark=triangle*,mark size=1.1pt,mark options={solid},const plot,mark repeat=4] coordinates {(0,0.9635) (1,0.9642) (2,1.2065) (3,1.2065) (4,1.2065) (5,1.2065) (6,1.2338) (7,1.2338) (8,1.2489) (9,1.2489) (10,1.2489) (11,1.2584) (12,1.2769) (13,1.2769) (14,1.2769) (15,1.2775) (16,1.2775) (17,1.2803) (18,1.2803) (19,1.2811) (20,1.2853) (21,1.2944) (22,1.3012) (23,1.3454) (24,1.3454) (25,1.3454)};
\addplot[black!55,dashdotted,semithick,mark=o,mark size=1pt,mark options={solid},const plot,mark repeat=4] coordinates {(0,0.9635) (1,0.9635) (2,0.9935) (3,0.9935) (4,0.9935) (5,1.1943) (6,1.2584) (7,1.2584) (8,1.2584) (9,1.2584) (10,1.2584) (11,1.2584) (12,1.2584) (13,1.2584) (14,1.2584) (15,1.2584) (16,1.2584) (17,1.2584) (18,1.2584) (19,1.2724) (20,1.2724) (21,1.2724) (22,1.2724) (23,1.2724) (24,1.2724) (25,1.2808)};
\legend{Full,Plan,Guide,Record}\end{axis}\end{tikzpicture}
\caption{Planning raises the peak; feedback speeds convergence.}
\Description{One step plot of best IPC so far versus recorded design index with four curves. Plan exceeds Record, Full exceeds Guide, Guide exceeds Record, and Full slightly exceeds Plan while reaching IPC 1.3 at its second edit instead of its ninth.}
\label{fig:traj}
\end{figure}
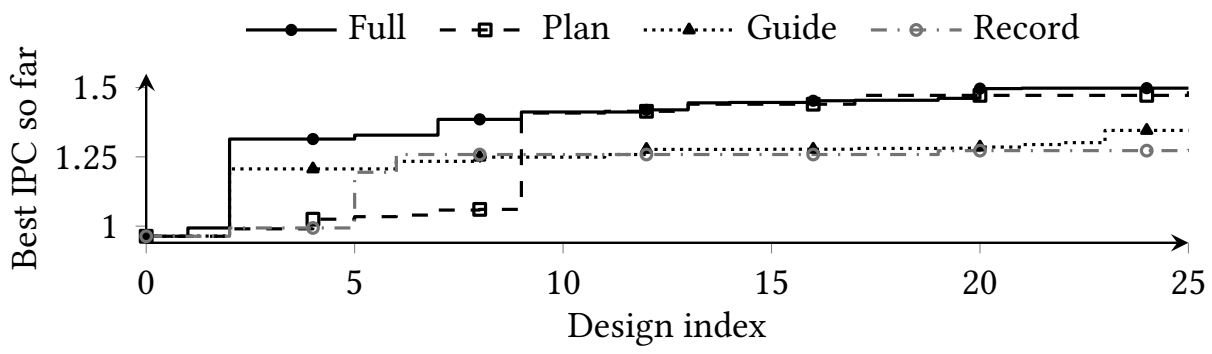

\begin{figure*}[t]
\centering
\begin{minipage}[b]{.40\textwidth}\centering
\begin{tikzpicture}\begin{axis}[
width=\linewidth,height=32mm,ymin=0,ymax=32,xmin=-.8,xmax=16.8,
xtick={0,8,16},ytick={0,10,20,30},
xlabel={Target support measurements},ylabel={IPC MAPE (\%)},
label style={font=\footnotesize},tick label style={font=\footnotesize},
axis lines=left,grid=none,
legend style={font=\scriptsize,draw=none,fill=none,at={(.5,1.08)},anchor=south,
legend columns=2,column sep=1pt,/tikz/every even column/.append style={column sep=3pt}}]
\addplot[black,densely dashed,mark=asterisk,mark size=1.4pt,mark options={solid},
error bars/.cd,y dir=both,y explicit,error bar style={line width=.4pt}]
table[x=budget,y=error,y error minus=lower,y error plus=upper,row sep=\\]{
budget error lower upper\\
0 9.00344207 0.00000000 0.00000000\\
8 9.00344207 0.00000000 0.00000000\\
16 9.00344207 0.00000000 0.00000000\\
};
\addlegendentry{Source IPC}
\addplot[black,densely dotted,mark=square,mark size=1.4pt,mark options={solid},
error bars/.cd,y dir=both,y explicit,error bar style={line width=.4pt}]
table[x=budget,y=error,y error minus=lower,y error plus=upper,row sep=\\]{
budget error lower upper\\
0 9.10852820 0.01337254 0.01874873\\
8 9.10852820 0.01337254 0.01874873\\
16 9.10852820 0.01337254 0.01874873\\
};
\addlegendentry{NN32 (source)}
\addplot[black,dashdotted,mark=triangle,mark size=1.4pt,mark options={solid},
error bars/.cd,y dir=both,y explicit,error bar style={line width=.4pt}]
table[x=budget,y=error,y error minus=lower,y error plus=upper,row sep=\\]{
budget error lower upper\\
0 9.13295376 0.02601439 0.04143693\\
8 18.12666921 7.86599535 12.19954250\\
16 6.03581646 3.13011137 3.03551753\\
};
\addlegendentry{NN8 (retrained)}
\addplot[black,dashed,mark=triangle*,mark size=1.4pt,mark options={solid},
error bars/.cd,y dir=both,y explicit,error bar style={line width=.4pt}]
table[x=budget,y=error,y error minus=lower,y error plus=upper,row sep=\\]{
budget error lower upper\\
0 9.10852820 0.01337254 0.01874873\\
8 15.02690562 2.40085346 3.21736075\\
16 7.22492493 1.88952201 2.37678285\\
};
\addlegendentry{NN32 (retrained)}
\addplot[black,dashdotdotted,mark=diamond,mark size=1.4pt,mark options={solid},
error bars/.cd,y dir=both,y explicit,error bar style={line width=.4pt}]
table[x=budget,y=error,y error minus=lower,y error plus=upper,row sep=\\]{
budget error lower upper\\
0 11.77177578 0.00000000 0.00000000\\
8 7.36487680 0.00000000 0.00000000\\
16 4.56318675 0.00000000 0.00000000\\
};
\addlegendentry{RF (retrained)}
\addplot[black,solid,thick,mark=*,mark size=1.4pt,mark options={solid},
error bars/.cd,y dir=both,y explicit,error bar style={line width=.4pt}]
table[x=budget,y=error,y error minus=lower,y error plus=upper,row sep=\\]{
budget error lower upper\\
0 6.33503467 1.53649357 1.53649357\\
8 1.80146116 0.17722091 0.17722091\\
16 0.87402917 0.07944428 0.07944428\\
};
\addlegendentry{DELTA}
\end{axis}\end{tikzpicture}
\par\small (a) All eight workloads
\end{minipage}\hfill
\begin{minipage}[b]{.30\textwidth}\centering
\begin{tikzpicture}\begin{axis}[
width=\linewidth,height=32mm,ybar=0pt,bar width=5pt,ymin=0,ymax=36,
xmin=.5,xmax=2.5,xtick={1,2},xticklabels={{Independent\\ALU},{Predictable\\branch}},
ytick={0,10,20,30},ylabel={IPC MAPE (\%)},
x tick label style={align=center},label style={font=\footnotesize},tick label style={font=\footnotesize},
axis lines=left,grid=none,
legend style={font=\scriptsize,draw=none,fill=none,at={(.5,1.08)},anchor=south,legend columns=2},
legend image code/.code={\draw[#1] (0cm,-.05cm) rectangle (.2cm,.1cm);}]
\addplot[draw=black,fill=white] coordinates {(1,33.32519829) (2,22.69473157)};
\addlegendentry{Source IPC}
\addplot[draw=black,fill=white,pattern=north east lines,pattern color=black] coordinates {(1,33.33986391) (2,22.70704829)};
\addlegendentry{NN32}
\addplot[draw=black,fill=black!50] coordinates {(1,33.45433605) (2,23.09716385)};
\addlegendentry{RF}
\addplot[draw=black,fill=black] coordinates {(1,12.77199076) (2,12.40215317)};
\addlegendentry{DELTA}
\end{axis}\end{tikzpicture}
\par\vspace{1.5mm}\small (b) No target support
\end{minipage}\hfill
\begin{minipage}[b]{.28\textwidth}\centering
\begin{tikzpicture}\begin{axis}[
width=\linewidth,height=32mm,ybar=0pt,bar width=10pt,ymin=-10,ymax=12,
xmin=.5,xmax=2.5,xtick={1,2},xticklabels={LocalBP,TAGE},ytick={-10,-5,0,5,10},
ylabel={IPC change (\%)},label style={font=\footnotesize},tick label style={font=\footnotesize},
axis lines=left,grid=none,
legend style={font=\scriptsize,draw=none,fill=none,at={(.5,1.08)},anchor=south,legend columns=2},
legend image code/.code={\draw[#1] (0cm,-.05cm) rectangle (.2cm,.1cm);}]
\addplot[draw=black,fill=white] coordinates {(1,-2.07305993) (2,10.72766640)};
\addlegendentry{ROB 64}
\addplot[draw=black,fill=black!50] coordinates {(1,-8.78057944) (2,6.04246365)};
\addlegendentry{ROB 192}
\draw[black,thin] (axis cs:.5,0)--(axis cs:2.5,0);
\end{axis}\end{tikzpicture}
\par\vspace{4mm}\small (c) Width $4\rightarrow8$
\end{minipage}
\caption{Width transfer: (a) 48 queries, ranges over two DELTA repetitions/five NN seeds; (b) no target support, DELTA rightmost; (c) random branch IPC changes from four-wide parents, averaged over L1D sizes. Lower error is better. Memory-based prediction transfers to a wider core, the gap is largest where width matters, and the same widening helps under TAGE but hurts under LocalBP.}
\Description{Three panels show prediction error over target support measurements, zero-support errors for independent additions and predictable branches, and opposite effects of pipeline widening under LocalBP and TAGE at two reorder-buffer sizes.}
\label{fig:transfer}
\end{figure*}
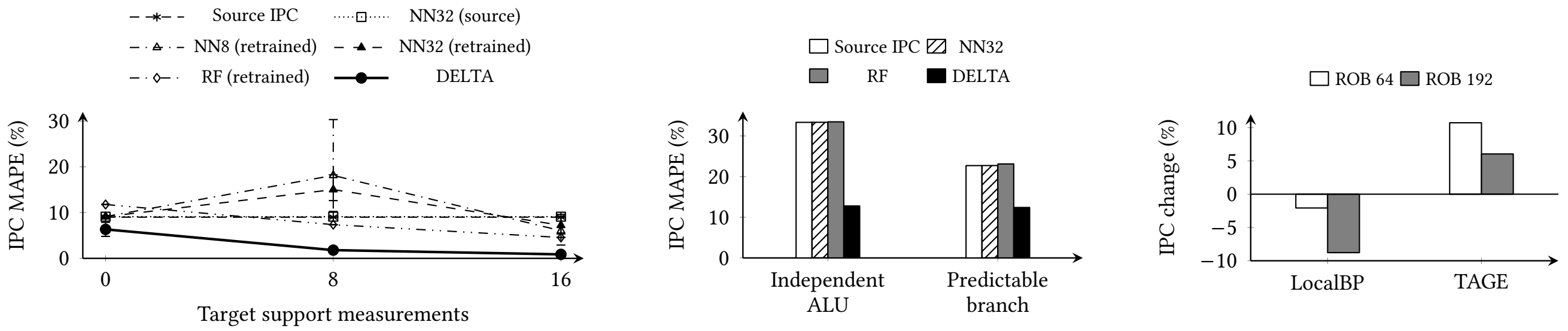

\subsection{Does elimination expose new root causes on RISC-V?}
\label{sec:riscv}
Algorithm~\ref{alg:explore} is applied unchanged to the RISC-V core with SonicBOOM Mega sizes (ROB 128, load/store queue (LSQ) 32, L1 stride, L2 BOP) on workloads outside the recorded histories. Hierarchical Navigable Small World (HNSW) vector search~\cite{hnsw} incurs 535,405 L1D misses in 20 million instructions, so the candidates are window, request, and cache capacity and prefetch coverage. First, single edits eliminate capacity: ROB 256, 16 MSHRs, and 1\,MiB L2 change cycles by $+0.02\%$, $-0.13\%$, and $+1.81\%$. Second, coverage remains although stride and BOP are present, so the missed addresses follow neither a constant stride nor a fixed offset. Third, the miss PCs locate the pattern: the search loop reads a neighbor list sequentially and loads $\mathit{base}+396\cdot\mathit{id}$ for each identifier; an element holds 132\,B of links, a 256\,B vector, and an 8\,B label. Fourth, a necessary condition separates this form from indirect prefetchers. The Indirect Memory Prefetcher (IMP)~\cite{imp} and the Differential-Matching Prefetcher (DMP)~\cite{dmp} prefetch \texttt{A[B[i]]} as $\mathit{base}+(\mathit{index}\ll s)$, so the scale is a power of two; DMP detects the pair by matching differences of target addresses against shifted differences of index values. Because $396=4\cdot 99$ is not a power of two, no shift reproduces the measured differences. Fifth, the separating control confirms it: with the same table, prefetch distance, and L1 placement, matching restricted to shifts 0 to 3 (scales 1 to 8) learns only one pattern, of scale 2, and gains 0.66\% over no L1 prefetcher.

The repair keeps differential matching and replaces the shift test with a division: $\mathit{scale}=\Delta\mathit{miss}/\Delta\mathit{index}$, accepted when two successive index pairs give the same base and scale. It learns scale 396 at three load PCs, learns how many lines beyond the target the loop reads (the vector), and prefetches the targets of the next three identifiers in the line already loaded. Memory reads fall 310,229$\to$179,016. The repair gains 7.32\% over no L1 prefetcher, 12.25\% over the base, and 18.23\% over the base when combined with ROB 256 and LSQ 64. Across nine matched configurations varying window, bus width, L2 prefetcher, prefetch distance, and stride, shift matching gains 0.66--1.32\% and the repair 4.47--14.98\%. Matched configurations also expose a sign reversal: removing stride gains 4.59\% with BOP and loses 16.9\% without it. Elimination thus returns new dependencies and a repair on another instruction set, core model, and workload class.

\section{Reuse for Search and Prediction}
\label{sec:baselines}\label{sec:search}\label{app:search}
The same records that support diagnosis also guide subsequent proposals and predictions. We ask how effectively DELTA searches, how accurately it predicts the next edit, and whether prediction transfers to a wider pipeline without retraining.

\subsection{How effectively does DELTA guide search?}
Under the reported policies and initial information, DELTA reaches the largest mean gain; planning improves peak IPC, while feedback reduces evaluations to the peak.

Over ten seeds with a shared checker, mean IPC gains at ten iterations are 46\% for DELTA, 36\% for GA, 33\% for Top-Down Greedy, and 19\% for Random. DELTA reaches IPC 1.44 in 20 evaluations versus over 60 for GA; Random does not reach it within 200, and Greedy does not reach it over its permitted edits. Greedy improves from 33\% at ten iterations to 35\% at twenty. Evaluation budgets count completed simulations rather than iterations.

Comparing the four variants isolates planning and bottleneck feedback (Figure~\ref{fig:traj}). Planning raises peak IPC by 16.08\% without feedback and 11.33\% with it. Feedback raises the peak by 5.04\% without planning and 0.75\% with planning, but Full peaks at edit 20 versus Plan's 24. Counting the baseline and stopping at the first peak, Record, Guide, Plan, and Full reach IPC 1.28081, 1.34536, 1.48671, and 1.49785 (gains 32.93\%, 39.63\%, 54.30\%, and 55.46\%) in 26, 24, 26, and 22 evaluations, or 1.27\%, 1.65\%, 2.09\%, and 2.52\% per evaluation, at model costs of \$1.57, \$1.51, \$1.80, and \$1.56. Planning improves the best design; feedback reduces the evaluations to reach it.

Full crosses IPC 1.31 with SlimAMPM at edit 2, whereas Plan reaches it with DCPT at edit 9. These first L2 prefetchers supply 60.0\% and 66.9\% of their gains. Record enables ISB at edit 5 and plateaus near 1.28; Guide plateaus near 1.35.

Across 25 workloads, harmonic mean IPC rises from 0.4828 to 1.0813 with GPT-5 ($2.24\times$) and 0.9733 with gpt-oss-120b ($2.02\times$). Geometric means of per-workload ratios are $1.40\times$ and $1.30\times$. Three memory-intensive workloads starting below IPC 0.17 exceed $3.7\times$ with both models; GPT-5 loses at most 1\% on any workload.

\subsection{How accurately can DELTA predict the next edit?}
\label{sec:surrogate}\label{app:surrogate}
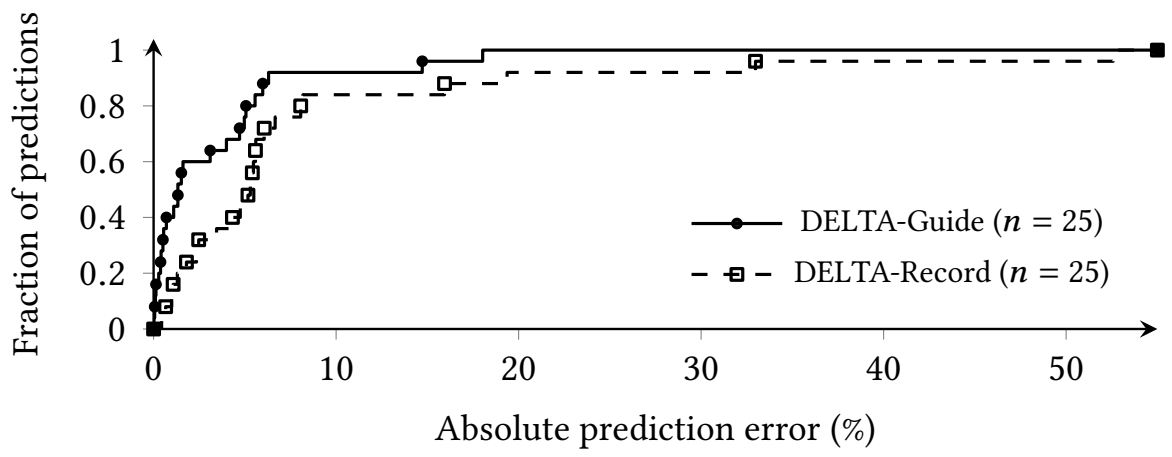
\begin{figure}[t]
\centering
\begin{tikzpicture}\begin{axis}[
width=.97\linewidth,height=35mm,xmin=0,xmax=55,ymin=0,ymax=1.04,
xlabel={Absolute prediction error (\%)},ylabel={Fraction of predictions},
label style={font=\footnotesize},tick label style={font=\footnotesize},
axis line style={black,line width=.45pt},grid=none,
legend style={font=\scriptsize,draw=none,fill=white,fill opacity=.85,text opacity=1,at={(.98,.06)},anchor=south east,legend columns=1}]
\addplot[black,solid,semithick,const plot,mark=*,mark size=.8pt,mark repeat=2] coordinates {(0.0000,0.0000) (0.0200,0.0400) (0.0600,0.0800) (0.1000,0.1200) (0.1300,0.1600) (0.2800,0.2000) (0.3800,0.2400) (0.4100,0.2800) (0.5100,0.3200) (0.5500,0.3600) (0.7000,0.4000) (1.1000,0.4400) (1.3300,0.4800) (1.3300,0.5200) (1.5200,0.5600) (1.6100,0.6000) (3.1000,0.6400) (3.9900,0.6800) (4.7100,0.7200) (4.9700,0.7600) (5.0600,0.8000) (5.5600,0.8400) (5.9800,0.8800) (6.3000,0.9200) (14.7200,0.9600) (18.0300,1.0000) (55.0000,1.0000)};
\addplot[black,dashed,semithick,const plot,mark=square,mark size=1.1pt,mark repeat=2,mark options={solid,fill=white}] coordinates {(0.0000,0.0000) (0.4400,0.0400) (0.6600,0.0800) (0.9900,0.1200) (1.0700,0.1600) (1.3000,0.2000) (1.8000,0.2400) (2.3500,0.2800) (2.4700,0.3200) (3.4500,0.3600) (4.3200,0.4000) (4.7500,0.4400) (5.1700,0.4800) (5.3000,0.5200) (5.4100,0.5600) (5.4700,0.6000) (5.5800,0.6400) (5.6000,0.6800) (6.0600,0.7200) (6.6600,0.7600) (8.0600,0.8000) (8.1800,0.8400) (15.9400,0.8800) (19.3700,0.9200) (32.9800,0.9600) (52.5800,1.0000) (55.0000,1.0000)};
\legend{DELTA-Guide ($n=25$),DELTA-Record ($n=25$)}
\end{axis}\end{tikzpicture}
\caption{IPC prediction error over 25 predictions per variant. Low error without offline training makes the predicted IPC a usable guide for the next edit.}
\Description{Empirical cumulative distributions of absolute percentage prediction error for DELTA-Guide and DELTA-Record.}
\label{fig:err}
\end{figure}

Guide predicts child IPC with 3.30\% MAPE and 1.33\% median error, using no offline simulation corpus (Figure~\ref{fig:err}). Fifteen of its 25 predictions are within 2\% and 23 within 6.3\%. Full, Plan, and Record have 3.36\%, 4.97\%, and 8.24\% MAPE; Record's median is 5.30\%. Guide's error falls from 4.96\% over its first 13 steps to 1.49\% over the last 12; Record's falls from 11.37\% to 4.84\%.

These histories contain different queries. Guide's initial predictor and L2 BOP edits have its largest plotted errors, 18.03\% and 14.72\%. Record:16 predicts ISB behind a 32\,KB L1I with 19.37\% error and a $-0.01\%$ measured response, while the same policy gains 20.21\% behind 64\,KB. Parent context therefore accompanies prediction feedback. Across Record, Plan, and Full, 12 TAGE-SC-L proposals follow an earlier regression; ten are regressive and provide ten of the 13 regressive contexts, and the 12 cost roughly 1{,}440 core hours.

\subsection{Can prediction transfer without retraining?}
\label{sec:transfer}
DELTA incorporates measurements through memory without supervised training, backpropagation, or fine-tuning. The study with fixed queries tests reuse across four-wide and eight-wide cores (Figure~\ref{fig:transfer}). With no target measurements, DELTA's IPC error is 6.34\%, versus 9.11\% for frozen NN32 and 9.00\% for source IPC reuse. With 8/16 support measurements, DELTA reaches 1.80\%/0.87\%, retrained NN8 18.13\%/6.04\%, and RF 7.36\%/4.56\%.

Independent additions expose the width response: source IPC reuse has 33.33\% error, NN32 33.34\%, and DELTA 12.77\% without target support. For predictable branches the corresponding errors are 22.69\%, 22.71\%, and 12.40\%. Retaining the source architecture and measured behavior helps distinguish additional execution bandwidth from the reversal across predictors of Section~\ref{sec:recovery}.

\section{Related Work}
\label{sec:background}
\paragraph{Pipeline slots and bottleneck analysis.}
Top-Down~\cite{yasin_ispass14}, implemented in gem5~\cite{ferrer_jsc25} and Icicle on RISC-V~\cite{icicle_iiswc25}, classifies slots as Retiring, Bad Speculation, Frontend Bound, or Backend Bound. DELTA pairs slot and IPC changes.

Critical-path prediction identifies instructions determining execution time~\cite{fields_isca01}; interaction cost analysis compares overlapping event latencies with individual contributions~\cite{fields_micro03}. ArchExplorer allocates resources to bottlenecks~\cite{archexplorer_micro23}; Gus perturbs modeled resources for instruction-level timing sensitivity~\cite{gus24}; CacheMind analyzes caches through traces and program semantics~\cite{cachemind_asplos26}. DELTA compares identical edits across parents, using IPC and counters to identify core, predictor, window, and cache interactions.

\paragraph{Design-space exploration.}
To reduce simulation cost, \citet{lee_hpca07} fit performance/power regression models to sampled simulations; \citet{ipek_asplos06} train neural networks for parameter tradeoffs. \citet{dubach_micro07} reuse simulated programs to predict new programs' design spaces with few simulations; \citet{li_dac16} combine sampling and boosting. ArchRanker~\cite{archranker_isca14} ranks candidate pairs. Bayesian optimization balances quality and uncertainty for accelerator design~\cite{reagen_islped17}; BOOM-Explorer~\cite{boomexplorer_iccad21} combines microarchitecture-aware active learning and Gaussian processes for RISC-V BOOM. ArchGym~\cite{archgym_isca23} supplies a common search--simulator interface. Beyond selecting designs, DELTA retains parents, edits, IPC, and counters for neutral and regressive steps.

\paragraph{Analytical and learned performance models.}
First-order superscalar~\cite{karkhanis_sigarch04} and interval~\cite{eyerman_tocs09} models add miss/misprediction penalties to steady-state throughput, modeling structures and overlap without training. Learned models fit labeled relationships. Concorde~\cite{concorde_isca25} combines analytical throughput bounds with training on trace features. SimNet~\cite{simnet_pomacs22} trains instruction latency predictors on simulator traces for GPU-accelerated simulation. TAO~\cite{tao_pomacs24} transfers via functional traces and an embedding shared across microarchitectures. Ithemal~\cite{ithemal_icml19} trains networks on measured timings for basic-block throughput; DiffTune~\cite{difftune_micro20} learns a differentiable simulator surrogate to fit simulator parameters. DELTA conditions fixed pretrained weights on run records to predict child IPC. New measurements enter memory without task-specific supervised training, backpropagation, or fine-tuning; retained parents and edits support prediction and root-cause elimination.

\paragraph{Search with language models.}
FunSearch~\cite{funsearch_nature24} and AlphaEvolve~\cite{alphaevolve25} use automated evaluation scores to guide search over generated programs. DELTA retains every evaluated edit's parent and response, including neutral and regressive outcomes.

\paragraph{Prefetching.}
Feedback control~\cite{fdp_hpca07} adjusts prefetching using accuracy, timeliness, and pollution. IMP~\cite{imp} and DMP~\cite{dmp} prefetch indirect accesses at power-of-two scales; elimination on HNSW locates a scale of 396 and repairs matching by division (Section~\ref{sec:riscv}).

\section{Conclusion}
\label{sec:conclusion}
DELTA makes measured design changes reusable for architectural reasoning. Parent alignment reveals when instruction supply enables prefetching, prediction changes the return from window capacity, and recovery or request coverage limits throughput. Successive counter checks, including IPC, connect these interactions to measured controls and controller repairs on x86 and RISC-V cores. The same records guide search and prediction without task-specific weight training, preserving useful designs and the comparisons that explain them.

\begin{acks}
We used Perplexity Computer to assist with drafting and restructuring scientific prose and organizing counter-based analyses. This assistance was distinct from the language-model components evaluated as part of DELTA. The authors take full responsibility for the paper's technical content, results, citations, and conclusions.
\end{acks}

\clearpage

\end{document}